\documentclass[10pt,aps,preprintnumbers,prd,noshowpacs,nofootinbib,noshowkeys,floatfix,superscriptaddress]{revtex4}
\usepackage[dvips]{graphics,graphicx}
\usepackage[colorlinks=true,linktocpage=true,linkcolor=blue,citecolor=blue]{hyperref}
\usepackage[usenames,dvipsnames]{color}
\usepackage{amsmath, amssymb,oldgerm}
\usepackage{multirow}
\usepackage{xcolor}
\usepackage[normalem]{ulem}  
\usepackage{braket}
\usepackage{slashed}
\usepackage[mathscr]{eucal}
\usepackage[dvips]{graphics,graphicx}
 \usepackage{caption}
\usepackage{subcaption}
\usepackage[version=4]{mhchem}

\newcommand{\n}{\nonumber}

\newcommand{\A}{\mathcal{A}}
\newcommand{\mC}{\mathfrak{C}}

\newcommand{\ms}{\mathfrak{s}}

\newcommand{\beq}{\begin{equation}}
\newcommand{\eeq}{\end{equation}}

\newcommand{\eqs}{Eqs.~}
\newcommand{\eq}{Eq.~}

\newcommand{\bbb}{\textcolor{black}}

\newcommand{\p}{\mathbf{p}}

\newcommand{\mL}{\mathcal{L}}

\newcommand{\ppartial}{\boldsymbol{\partial}}
\newcommand{\vb}{\mathbf{v}}

\begin{document}

\title{Spin kinetic theory with a nonlocal relaxation time approximation}

\author{Nora Weickgenannt}

\affiliation{Institut de Physique Th\'eorique, Universit\'e Paris Saclay, CEA, CNRS, F-91191 Gif-sur-Yvette, France}

\author{Jean-Paul Blaizot}

\affiliation{Institut de Physique Th\'eorique, Universit\'e Paris Saclay, CEA, CNRS, F-91191 Gif-sur-Yvette, France}

\begin{abstract}
 We present a novel relaxation time approximation for kinetic theory with spin which takes into account the nonlocality of particle collisions. In particular, it models the property of the microscopic nonlocal collision term to vanish in global, but not in local equilibrium. We study the asymptotic distribution function obtained as the solution of the Boltzmann equation within the nonlocal relaxation time approximation in the limit of small gradients and short relaxation time. We show that the resulting polarization agrees with the one obtained from the Zubarev formalism for a certain value of a coefficient that determines the time scale on which orbital angular momentum is converted into spin. This coefficient can be identified with a parameter related to the pseudo-gauge choice in the Zubarev formalism. Finally, we demonstrate how the nonlocal collision term generates polarization from vorticity by studying a nonrelativistic rotating cylinder both from kinetic and hydrodynamic approaches, which are shown  to be equivalent in this example.
\end{abstract}

\maketitle

\section{Introduction}

The interplay between polarization and orbital angular momentum is a phenomenon that leads to a wide range of interesting effects. While the total angular momentum is always conserved, its splitting into orbital angular momentum and spin may vary with time, leading to the possibility that  polarization be generated from rotation. A well-known nonrelativistic example for the conversion of orbital angular momentum from the rotation of the medium into particle polarization is the Barnett effect~\cite{Barnett:1935}. Similar effects have been studied in the context of nonrelativistic fluids with internal angular momentum such as micropolar fluids~\cite{Lukaszewicz1999} or chiral active fluids~\cite{banerjee2017odd}, as well as in spintronics~\cite{takahashi2016spin}. A relativistic manifestation of an analogous phenomenon has been identified in heavy-ion collisions, where Lambda hyperons get polarized along the global angular momentum~\cite{Liang:2004ph,Voloshin:2004ha,Betz:2007kg,Becattini:2007sr}.  In Ref.~\cite{Becattini:2017gcx}, a momentum-dependent polarization along the beam direction (``local Lambda polarization") originating from local vortices has been predicted, a phenomenon which is still not completely understood~\cite{Adam:2018ivw,ALICE:2019aid,STAR:2019erd,Florkowski:2019qdp,Florkowski:2019voj,Becattini:2020ngo,Banerjee:2024xnd}. While the first theoretical calculations~\cite{Becattini:2013fla,Becattini:2017gcx} turned out to be inconsistent with the experimental data~\cite{Adam:2018ivw,ALICE:2019aid,STAR:2019erd}, recent progress has been made in Refs.~\cite{Liu:2021uhn,Fu:2021pok,Becattini:2021suc,Becattini:2021iol}. In particular, the authors of Refs.~\cite{Becattini:2021suc,Becattini:2021iol}   derive the polarization in local equilibrium using the Zubarev formalism,  and find that the contributions from thermal shear may explain the measured momentum dependence. However, the role of nonequilibrium effects is left aside in this approach. Furthermore, the results obained in the Zubarev formalism depend on the choice of a pseudo-gauge~\cite{Buzzegoli:2021wlg}, which obscures the interpretation of the results. It thus appears desirable to have an alternative description, which remains valid  beyond local equilibrium and is independent of the pseudo-gauge. Such a theory can be provided, e.g., through spin kinetic theory and spin hydrodynamics.

In the past years, relativistic kinetic theory with spin has been extensively studied~\cite{Weickgenannt:2019dks,Gao:2019znl,Hattori:2019ahi,Wang:2019moi,Weickgenannt:2020aaf,Liu:2020flb,Weickgenannt:2021cuo,Sheng:2021kfc,Wagner:2022amr,Wagner:2023cct,Fang:2024vds,Wang:2024lis}. As already outlined in an old seminal paper for the nonrelativistic case~\cite{hess1966kinetic} and recently demonstrated for the relativistic case in Refs.~\cite{Weickgenannt:2020aaf,Weickgenannt:2021cuo}, where the nonlocal collision term for spin-1/2 particles was obtained from the Wigner-function formalism, it is essential to take into account the nonlocality of particle collisions to describe the conversion between orbital angular momentum and spin in kinetic theory: the nonlocal part of the collision term is responsible for aligning spin polarization along the vorticity direction, while the local part drives spin diffusion.\footnote{Spin diffusion here refers to the process of eliminating a possible initial inhomogeneous polarization through the transport of spin-carrying particles, while the spin itself is conserved in particle collisions.} Due to the complexity of solving kinetic theory directly, it is often convenient to study instead  equations of motion for spin hydrodynamics~\cite{Florkowski:2017ruc,Florkowski:2017dyn,Florkowski:2018fap,Montenegro:2018bcf,Hattori:2019lfp,Bhadury:2020puc,Singh:2020rht,Montenegro:2020paq,Gallegos:2021bzp,Fukushima:2020ucl,Li:2020eon,Wang:2021ngp,Hu:2021pwh,Hongo:2021ona,Daher:2022xon,Weickgenannt:2022zxs,Weickgenannt:2022jes,Gallegos:2022jow,Cao:2022aku,Weickgenannt:2022qvh,Biswas:2023qsw,Weickgenannt:2023btk,Weickgenannt:2023bss,Daher:2024bah,Wagner:2024fhf,Drogosz:2024gzv,Wagner:2024fry}, which may be obtained, e.g., from the Boltzmann equation.
 In Ref.~\cite{Weickgenannt:2022zxs} second-order spin hydrodynamics was derived from nonlocal spin kinetic theory. It was found that the nonlocal collision term contributes new terms to the polarization vector in the Navier-Stokes limit, which are proportional to both the difference between the thermal vorticity and the spin potential, and to the thermal shear. However, a numerical implementation of the equations of motion of Ref.~\cite{Weickgenannt:2022zxs} is still missing and may be demanding due to the presence of various complicated collision integrals. Thus, for practical applications both of spin kinetic theory and hydrodynamics it may be useful to provide a simple approximation of the nonlocal collision term, which models its essential microscopic properties. 
 
In this paper, we suggest a nonlocal relaxation time approximation (NLRTA). The conventional relaxation time approximation (RTA) is a widely and successfully used model for the local collision term, whose main role is to drive the system to local equilibrium. However, the nonlocal collision term does not drive the system to local equilibrium, since in general it does not vanish under local-equilibrium conditions~\cite{Weickgenannt:2020aaf,Weickgenannt:2021cuo}. In our NLRTA, we mimic this property by adding to the usual RTA a term of first order in an expansion in powers of $\hbar$. This term is chosen so as to vanish under the conditions of global equilibrium, in particular when the polarization is determined by the vorticity. Furthermore, we show that there exists another state for which the NLRTA collision term vanishes. We refer to  the corresponding distribution function as to the asymptotic distribution function. It is approached in the limit of small gradients and short relaxation time. It turns out that the polarization obtained from this asymptotic distribution function coincides with that obtained from the Zubarev formalism mentioned above. Interestingly, a parameter related to the pseudo-gauge choice in the Zubarev formalism can be identified with a coefficient in the NLRTA corresponding to the ratio of the relaxation time for spin diffusion to the time scale associated to the conversion between vorticity and polarization. This suggests an interpretation of the pseudo-gauge choice in the Zubarev formalism as a way to model the effectiveness of the conversion between orbital angular momentum and spin. 

To investigate how polarization emerges from rotation, we study the example of a nonrelativistic rigidly rotating cylinder. For this simple example, the Boltzmann equation  can be solved exactly within the NLRTA. We find that when the system has initially zero polarization but nonzero vorticity, a polarization emerges through the nonlocal part of the collision term, and the approach to the final form of the polarization is exponential. Furthermore, we show that for this example one can also obtain the exact solution of the Boltzmann equation without solving the latter microscopically, but instead by solving hydrodynamic equations of motion for micropolar fluids.

This paper is organized as follows. In Sec.~\ref{reviewsec} we review the basic concepts of spin kinetic theory with a nonlocal collision term. Then, in Sec.~\ref{nlrtasec}, we introduce the NLRTA. In Sec.~\ref{fresec} we outline the expansion of the distribution function for short relaxation time and small gradients, the so-called fast-relaxation expansion. The polarization obtained at the leading order in this expansion is presented and compared to results for the polarization in heavy-ion collision obtained from the Zubarev formalism in Sec.~\ref{hicsec}. Section~\ref{cylsec} is dedicated to the example of a nonrelativistic rotating cylinder. Finally, we provide conclusions in Sec.~\ref{concsec}.
As usual in spin kinetic theory, we work up to first order in $\hbar$ throughout the paper. We use natural units, but keep $\hbar$ explicit in most places, since it serves as a power-counting parameter. Furthermore, we sum over repeated indices and use the following notations and conventions: $\tilde{A}^{\mu\nu}\equiv \epsilon^{\mu\nu\lambda\rho}A_{\lambda\rho}$, $a\cdot b\equiv a_\mu b^\mu$, $a_{[\mu}b_{\nu]}\equiv a_\mu b_\nu-a_\nu b_\mu$, $a_{(\mu}b_{\nu)}\equiv a_\mu b_\nu+a_\nu b_\mu$, $\mathbf{a}\cdot\mathbf{b}\equiv a^i b^i$, $\mathbf{a}_\bot\equiv (a^x,a^y,0)$,  $g_{\mu\nu}=\text{diag}(+,-,-,-)$, $\epsilon^{0123}=-\epsilon_{0123}=1$.

\section{Spin kinetic theory with a nonlocal collision term}
\label{reviewsec}

In this section, we shortly review the results of Refs.~\cite{Weickgenannt:2020aaf,Weickgenannt:2021cuo}, where a nonlocal collision term for massive spin-1/2 particles was derived. As already mentioned, including a nonlocality in the collision term is essential to convert orbital angular momentum into spin polarization. This can be understood as follows. In kinetic theory, the collisional invariants, charge, energy-momentum, and total angular momentum, are conserved in a  collision between two particles. If the collision term is local, in other words, if the particles collide with zero impact parameter, the orbital angular momentum before and after the collision vanishes and the spin angular momentum is separately conserved. In this case, the dynamics is determined purely by spin diffusion. On the other hand, with a nonlocal collision term, the particles collide with a finite, momentum-dependent, impact parameter. Hence, the orbital angular momentum before the collision differs from the one after the collision, and this difference is converted into spin. This process generates polarization from the vorticity of the fluid. 

The dynamics of a system of massive spinning particles in the kinetic regime can be described by a distribution function $f(x,p,\ms)$, which depends on the space-time $x$, the particle four-momentum $p$ and the particle spin four-vector $\ms$. \bbb{This distribution function is defined in the extended phase space $(x,p,\ms)$ and is directly related to components of the Wigner function. It was shown in Ref.~\cite{Weickgenannt:2020aaf} that the constraint equations for the Wigner function restrict the momentum to its mass-shell and the spin four vector to be orthogonal to momentum. Furthermore, the dynamical equations for the Wigner function yield the equation of motion for $f$. One finds that} its evolution is determined by the on-shell Boltzmann equation~\cite{Weickgenannt:2020aaf,Weickgenannt:2021cuo}
\begin{equation}
    p\cdot\partial f= \mC[f] \label{boltz}
\end{equation}
with
\begin{align}
{\mC}[f]   =& \int d\Gamma_1 d\Gamma_2 d\Gamma^\prime\,    {\mathcal{W}}\,  
[f(x+\Delta_1,p_1,\ms_1)f(x+\Delta_2,p_2,\ms_2)-f(x+\Delta,p,\ms)f(x+\Delta^\prime,p^\prime,\ms^\prime)]
\label{finalcollisionterm}
\end{align}
being the nonlocal collision term.  Here we defined $d\Gamma\equiv dP\, dS(p)$ 
as the integration measure over on-shell momentum space, $dP\equiv [1/(2\pi\hbar)^3] d^4p\, \delta(p^2-m^2)$, and
spin space, $dS(p)\equiv  (\sqrt{p^2}/{\sqrt{3}\pi})  d^4\ms\, \delta(\ms\cdot\ms+3)\delta(p\cdot \ms)$. Furthermore, $\mathcal{W}$ is the transition rate for the collision; see, e.g., Eq.\ (5) in
Ref.~\cite{Weickgenannt:2022zxs} for the the explicit expression. The space-time shifts $\Delta^\mu$ characterize the nonlocality of the collision measured  by the displacements of the incoming and outgoing particles. Assuming that the scattering amplitude is constant over the typical range  $\Delta$ of these displacements,  one obtains~\cite{Weickgenannt:2020aaf,Weickgenannt:2021cuo}
\begin{equation}
\Delta^\mu\equiv-\frac{\hbar}{2m (E_p+m)}\epsilon^{\mu\nu\alpha\beta}p_\nu t_\alpha \ms_\beta \; ,
\label{delta}
\end{equation}
where $E_p\equiv \sqrt{\p^2+m^2}$ is the particle energy and $t\equiv(1,\mathbf{0})$ is the time unit vector. The exact, rather complicated, form of $\Delta^\mu$ can be found in Ref.~\cite{Wagner:2022amr}. Note that $\Delta$ appears here naturally as being of order $\hbar$. \bbb{We remark that for the derivation of \eq\eqref{finalcollisionterm} one employs standard assumptions of kinetic theory, in particular a low-density approximation.}

As discussed in Refs.~\cite{Weickgenannt:2020aaf,Weickgenannt:2021cuo,Weickgenannt:2022zxs}, the nonlocal collision term features specific properties related to the notion of equilibrium. In spinless kinetic theory, local equilibrium is defined as a state where the collision term vanishes, while the left-hand side of the Boltzmann equation \eqref{boltz} is nonzero. On the other hand, in global equilibrium both sides of the Boltzmann equation vanish separately. The same concepts apply to spin kinetic theory if the collision term is local, i.e., if all shifts $\Delta^\mu$ in \eq\eqref{finalcollisionterm} are set to zero. In that case, the local-equilibrium distribution function becomes a function  of the sum of the collisional invariants\footnote{We assume here that the fluid is uncharged.} $p^\mu$ and $\Sigma_\ms^{\mu\nu}\equiv -(1/m)\epsilon^{\mu\nu\alpha\beta}p_\alpha\ms_\beta$, where the latter quantity is the so-called dipole-moment tensor which is conserved in local collisions, each contracted with a Lagrange multiplier. \bbb{Note that the whole formalism presented in this paper is valid only for massive particles.} Up to first order in $\hbar$, the local-equilibrium distribution function then reads  
\begin{equation}
f_\text{LE} =  e^{-\beta\cdot p} \left(1+\frac\hbar4 \Omega_{\mu\nu}\Sigma_\ms^{\mu\nu} \right) \; . 
\label{fle}
\end{equation}
The Lagrange multipliers $\beta^\mu$ and $\Omega^{\mu\nu}$ correspond to the ratio of the fluid velocity to the  temperature and to the spin potential, respectively. For the sake of simplicity, we neglect quantum statistics and use a simple Boltzmann distribution in \eq\eqref{fle}. However, including proper statistics  would not change any of our conclusions. It is important to note that, in general, the nonlocal collision term \eqref{finalcollisionterm} does not  vanish for the local-equilibrium distribution function \eqref{fle}, which contradicts the usual notion of local equilibrium. Typically, local equilibrium refers to situations where effects on the scale of the Knudsen number, defined as the ratio between  the mean free path and the scale of  hydrodynamic phenomena, are ignored.  However the nonlocality of the collision term is of the order of the ratio of the Compton wave length to the hydrodynamic scale. It is thus typically smaller than the Knudsen number, implying that collisions generate important dynamics not captured by the local equilibrium distribution function (\ref{fle}) (see Refs.~\cite{Weickgenannt:2021cuo,Weickgenannt:2022zxs} for detailed discussion of this issue). It has been shown from the explicit form of the nonlocal collision term in Ref.~\cite{Weickgenannt:2020aaf} that the latter vanishes if the distribution function assumes the form \eqref{fle} and if, in addition, the following conditions are fulfilled:
\begin{align}
\partial_{(\mu}\beta_{\nu)}&=0   \; ,\n\\
-\frac12\partial_{[\mu}\beta_{\nu]}&=\Omega_{\mu\nu} \; .
\label{globeq}
\end{align}
The conditions \eqref{globeq} imply that both sides of the Boltzmann equation \eqref{boltz} vanish separately, corresponding to a situation of global equilibrium. According to the second line in \eq\eqref{globeq}, the spin potential in global equilibrium is determined by the thermal vorticity $\varpi_{\mu\nu}\equiv-(1/2)\partial_{[\mu}\beta_{\nu]}$ showing that the fluid is polarized along the vorticity direction. Out of equilibrium, the spin potential is treated dynamically, and its equations of motion are determined from the conservation of the total angular momentum in a collision,
\begin{equation}
    \int d\Gamma\, \left( \Delta^{[\mu} p^{\nu]}+\frac\hbar2 \Sigma_\ms^{\mu\nu} \right) \mC[f]=0\; , \label{match}
\end{equation}
where the first term in the brackets constitutes the orbital angular momentum and the second term the spin.

\section{The nonlocal relaxation time approximation}
\label{nlrtasec}

Once the interaction is specified, the collision term \eqref{finalcollisionterm} can be computed exactly. However, solving kinetic theory or hydrodynamic equations of motion with the exact collision integrals  is  often highly involved, even in the case of a local collision term. A widely used approximation to model the exact local collision term  is the so-called relaxation time approximation (RTA),
\begin{equation}
    p\cdot\partial f=-p\cdot u \frac{f-f_\text{LE}}{\tau_R}\; , \label{classrta}
\end{equation}
where $\tau_R$ is the relaxation time. Just like the exact local collision term, the right-hand side of \eq\eqref{classrta} drives the system to local equilibrium, while the left-hand side drives it away from local equilibrium. If the gradients of the distribution function are much smaller than the inverse relaxation time, the dynamics is controlled by the right-hand side of \eq\eqref{classrta}, and the system  is then said to be in local equilibrium.

We cannot use \eq\eqref{classrta} to model the nonlocal collision term \eqref{finalcollisionterm} since, as we just argued, the latter does not drive the system to local equilibrium, unless the conditions \eqref{globeq} are fulfilled.  Instead, we search for an approximation of the collision term which reduces to \eq\eqref{classrta} at zeroth order, but has an additional contribution at first order whose role is to  model the properties of the nonlocal collision term. It is intuitive to choose this contribution to be proportional to $\Delta^\mu$. Furthermore, it should vanish when the global-equilibrium conditions \eqref{globeq} are fulfilled. Thus, we suggest the following nonlocal relaxation time approximation (NLRTA):
\begin{equation}
    p\cdot\partial f=-p\cdot u \frac{f-f_\text{LE}}{\tau_R}+ \xi\frac{p\cdot u}{\tau_R} \Delta^\mu p^\nu (\partial_\mu \beta_\nu+\Omega_{\mu\nu})f^{(0)}\; .
    \label{nlrta}
\end{equation}
Here we introduced a parameter $\xi$ which controls how fast the spin potential relaxes to thermal vorticity compared to the relaxation time $\tau_R$ for the diffusive process of relaxation to local equilibrium. While the first term in \eq\eqref{nlrta} corresponds to the local part of the collision term, which drives the system to local equilibrium through (spin) diffusion, the second term mimics the effect of the nonlocal part, which is responsible for aligning spin polarization with vorticity, cf.\ Ref.~\cite{Weickgenannt:2020aaf}. When the dynamics of the system is driven by the collision term, there will be a competition between these two effects. The parameter $\xi$ hence determines how effective the conversion of angular momentum is. In general, $\xi$ may depend on momentum.
Note that in the NLRTA \eqref{nlrta} we already neglected potential terms of second order in $\hbar$. The NLRTA is constructed such that it models the properties of the microscopic nonlocal collision term, i.e., its effect cannot be reduced to driving the distribution function to the local-equilibrium form \eqref{fle}. Instead, the distribution function may approach a different form at long time scales. This will be discussed in the next section. 

We close this section with a remark on the matching conditions. Using an RTA requires to choose a specific form of the matching conditions to be consistent with conservation laws. For instance, to conserve the four-momentum in a collision, one needs to fulfill
\begin{equation}
    \int d\Gamma\, p^\mu \mC[f]=0\; .
\end{equation}
Inserting the right-hand side of \eq\eqref{classrta} for the collision term, we find that we need to impose
\begin{equation}
    \int d\Gamma\, p\cdot u\,  p^\mu f= \int d\Gamma\, p\cdot u\, p^\mu f_\text{LE}\; ,
    \label{emmatch}
\end{equation}
which is known as the Landau matching condition. This relation determines the Lagrange multiplier $\beta^\mu$ in the local-equilibrium distribution function \eqref{fle}. Analogously, to preserve the conservation of total angular momentum \eqref{match} in the NLRTA \eqref{nlrta}, we require the matching condition
\begin{equation}
    \int d\Gamma\, \left( \Delta^{[\mu} p^{\nu]}+\frac\hbar2 \Sigma_\ms^{\mu\nu} \right) \left[p\cdot u \frac{f-f_\text{LE}}{\tau_R}- \xi\frac{p\cdot u}{\tau_R} \Delta^\mu p^\nu (\partial_\mu \beta_\nu+\Omega_{\mu\nu})f^{(0)} \right]=0\; ,
    \label{ammatch}
\end{equation}
which defines the spin potential. Note that $\Delta^\mu$ is of first order in $\hbar$, while $\Sigma_\ms^{\mu\nu}$ is of zeroth order, and $f$ and $f_\text{LE}$ contain terms both of zeroth and first order. However, zeroth-order terms in $f$ and $f_\text{LE}$ are independent of $\ms^\mu$, such that their contributions vanish when performing the spin integration, since the terms in the first round brackets are linear in $\ms^\mu$. 
\bbb{We remark that in principle, one could solve the Boltzmann equation with the NLRTA even without imposing any matching condition. However, in that case, the conservation laws would be violated and the Lagrange multipliers $\beta^\mu$ and $\Omega^{\mu\nu}$ would remain undetermined. Therefore, it is crucial to impose \eqs\eqref{emmatch} and \eqref{ammatch}.}

\section{Fast-relaxation expansion}
\label{fresec}

A commonly used method to derive hydrodynamics from kinetic theory is the gradient expansion around local equilibrium. In this section, we demonstrate how a similar expansion can be performed in spin kinetic theory within the NLRTA.
Assuming that the relaxation time $\tau_R$ is short and the gradients in the system are sufficiently small, we may perform an expansion around the distribution function which is a solution of \eq\eqref{nlrta} for $\tau_R\rightarrow 0$. We shall refer to this particular distribution function as to the asymptotic distribution function $f_\infty$. Note that $f_\infty\neq f_\text{LE}$ in general. 

The  expansion around the asympototic distribution is not a mere gradient expansion. Indeed, the second term on the right-hand side of \eq\eqref{nlrta} is of first order in gradients, but of zeroth order in $\tau_R$ after multiplying the whole equation with $\tau_R$. We shall therefore distinguish the expansion in $(\hbar/m)\partial$ (dubbed $\hbar$-expansion for simplicity),
\begin{equation}
    f=f^{(0)}+\hbar f^{(1)}+\mathcal{O}(\hbar^2)
    \label{hbarexp}
\end{equation}
and the expansion in $\tau_R\partial$, referred to as fast-relaxation expansion,
\begin{equation}
    f=f_\infty+\delta f+\mathcal{O}(\tau_R^2\partial^2)\; .
    \label{frexp}
\end{equation}
Each term on the right-hand side of \eq\eqref{frexp} can thus be separately expanded in $\hbar$. At zeroth order in $\hbar$, the asymptotic distribution function is given by the well-known local equilibrium distribution function
\begin{equation}    f_\infty^{(0)}=f^{(0)}_\text{LE}\; .
\end{equation}
Through the coefficient $\xi$ in \eq\eqref{nlrta}, we have an additional parameter to model different values of the time scale of angular-momentum conversion associated with the nonlocal collision term. Let us assume that $\xi\sim 1$ in the fast-relaxation expansion. We then find the following asymptotic distribution function as a solution of \eq\eqref{nlrta} for $\tau_R\rightarrow 0$,
\begin{equation}
\hbar f^{(1)}_\infty=    \hbar f^{(1)}_\text{LE}+\xi\Delta^\mu p^\nu (\Omega_{\mu\nu}+\partial_\mu\beta_\nu)f_\text{LE}^{(0)}\; . \label{finfty1}
\end{equation}
Therefore the asymptotic distribution function gains additional terms compared to the local-equilibrium distribution function, which are proportional to the difference between the spin potential and the thermal vorticity, and to the so-called thermal shear $(1/2)\partial_{(\mu}\beta_{\nu)}$. These terms are important for the polarization in heavy-ion collisions, as will be discussed in the next section.

We may obtain the distribution function at next-to-leading order by taking into account all terms of order $\tau_R$ in \eq\eqref{nlrta}. At zeroth order in $\hbar$ we obtain 
\begin{equation}
    p\cdot \partial f^{(0)}_\text{LE}=-\frac{p\cdot u}{\tau_R}\delta f^{(0)} \; , \label{gradex0}
\end{equation}
which corresponds to the usual first-order distribution function within a gradient expansion around local equilibrium. At first order in $\hbar$ we find
\begin{equation}
  \hbar  \tau_R p\cdot\partial f^{(1)}_\infty = -p\cdot u\, \hbar \delta f^{(1)}+ p\cdot u\, \xi\Delta^\mu p^\nu (\Omega_{\mu\nu}+\partial_\mu\beta_\nu)\delta f^{(0)}\; .
    \label{11}
\end{equation}
Inserting \eq\eqref{finfty1} into \eq\eqref{11} we obtain up to first order in $\hbar$
\begin{align}
  \delta f&=  -\frac{\tau_R}{p\cdot u} p\cdot\partial \left[ f_\text{LE}+\xi\Delta^\mu p^\nu (\Omega_{\mu\nu}+\partial_\mu\beta_\nu)f_\text{LE}^{(0)} \right]+ \xi\Delta^\mu p^\nu (\Omega_{\mu\nu}+\partial_\mu\beta_\nu)\delta f^{(0)}\n\\
 &=  \frac{\tau_R}{p\cdot u} \bigg[p^\lambda p^\rho(\partial_\lambda \beta_\rho)\left(1+\frac\hbar4 \Omega_{\mu\nu}\Sigma_\ms^{\mu\nu}\right)+ \frac\hbar4 p\cdot \partial \Omega_{\mu\nu}\Sigma_\ms^{\mu\nu}-\xi\Delta^\mu p^\nu p^\lambda (\partial_\lambda\Omega_{\mu\nu}+\partial_\mu\partial_\lambda\beta_\nu)\n\\
 &+2\xi \Delta^\mu p^\nu p^\lambda p^\rho (\Omega_{\mu\nu}+\partial_\mu\beta_\nu)\partial_\lambda \beta_\rho \bigg]f^{(0)}_\text{LE}\; ,
\end{align}
where we made use of \eqs\eqref{fle} and \eqref{gradex0} in the second step.
This distribution function can in principle be used to calculate spin-dependent quantities (e.g.\ the spin tensor) up to first order in the fast-relaxation expansion. In the nonrelativistic limit, to be discussed in Sec.~\ref{cylsec}, we show that this is possible without meeting conceptual problems. In the relativistic case,  the equations of motion for the spin potential derived within the first-order expansion would have to be checked for their stability and causality property  cf.\ Refs.~\cite{Bemfica:2017wps,Bemfica:2019knx,Kovtun:2019hdm,Bemfica:2020zjp,Hoult:2020eho,Weickgenannt:2023btk,Abboud:2023hos}. Such issues would require further analysis, and will not be addressed in the present paper.

\section{Asymptotic polarization in relativistic heavy-ion collisions}
\label{hicsec}

One of the main motivations to study relativistic spin kinetic theory and hydrodynamics is to investigate polarization phenomena in heavy-ion collisions. In this section, we compare the polarization obtained from the fast-relaxation expansion at leading order to the respective expressions derived in Refs.~\cite{Becattini:2021suc,Buzzegoli:2021wlg} from the so-called Zubarev approach. The latter, when applied to heavy-ion phenomenology,  appear to be consistent with experimental data.  However, the Zubarev approach relies on an explicit form of the energy-momentum and spin tensors and hence suffers from ambiguities emerging from the choice of pseudo gauge. While the choice of a pseudo gauge would not matter if one were doing an exact calculation, results obtained within approximations will most often be sensitive to the particular choice that is adopted. We will argue here that the pseudo-gauge freedom in the Zubarev formalism may be given a ``physical interpretation'' in terms of the typical time scale of the conversion between orbital angular momentum and spin.  

The momentum-dependent physical polarization vector can be obtained from the distribution function as~\cite{Weickgenannt:2020aaf}
\begin{equation}
    \A^\mu=\int dS(p)\, \ms^\mu f \; . \label{amu}
\end{equation}
If we use the local-equilibrium distribution function \eqref{fle} in \eq\eqref{amu}, we obtain
\begin{equation}
    \A^\mu_\text{LE}=-\frac{\hbar}{4m}\tilde{\Omega}^{\mu\nu}p_\nu f^{(0)}_\text{LE} \; .
    \label{ale}
\end{equation}
In global equilibrium with $\Omega^{\mu\nu}=\varpi^{\mu\nu}$, this agrees with the polarization vector derived with a thermodynamic approach in Ref.~\cite{Becattini:2013fla} [see also Refs.~\cite{Becattini:2013vja,Becattini:2015ska,Becattini:2016gvu,Karpenko:2016jyx,Pang:2016igs,Xie:2017upb} for related work], which is consistent with experimental data for the global Lambda polarization in heavy-ion collisions.

On the other hand, the local Lambda polarization cannot be adequately described by \eq\eqref{ale} with $\Omega^{\mu\nu}=\varpi^{\mu\nu}$~\cite{STAR:2019erd}. This has led to a large number of studies  in the last years. Beside kinetic theory, the main formalism used to derive the polarization vector in the context of heavy-ion collisions is the Zubarev formalism~\cite{zubarev1966statistical}. Including contributions from thermal shear within this approach, one finds that the polarization in thermodynamic local  equilibrium is given by~\cite{Becattini:2021suc} 
\begin{equation}
 \A^\mu_{Z} = -\frac{\hbar}{4m} \left[ \tilde{\varpi}^{\mu\nu}p_\nu+\frac{1}{E_p} \epsilon^{\mu\lambda\rho\sigma}p_\rho {t}_\sigma p^\nu\partial_{(\lambda} \beta_{\nu)}\right]  f^{(0)}_\text{LE} \; .
 \label{azub}
\end{equation}
Recent computations indicate that \eq\eqref{azub} may correctly describe the momentum dependence of Lambda polarization~\cite{Alzhrani:2022dpi,Jiang:2023vxp,Palermo:2024tza}.
Note that the definition of thermodynamic local equilibrium in Ref.~\cite{Becattini:2021suc} is different from the definition of kinetic local equilibrium, and \eq\eqref{azub} corresponds in the language of the present paper to the asymptotic polarization [and not the local-equilibrium polarization \eqref{ale}]. Inserting the distribution function \eqref{finfty1} into \eq\eqref{amu} and using the form of $\Delta^\mu$ in \eq\eqref{delta}, the asymptotic polarization is obtained as
\begin{equation}
    \A^\mu_\infty = -\frac{\hbar}{4m} \left[ \tilde{\Omega}^{\mu\nu}p_\nu+\frac{2}{E_p+m} \xi\epsilon^{\mu\lambda\rho\sigma}p_\rho {t}_\sigma p^\nu(\partial_\lambda \beta_\nu+\Omega_{\lambda\nu})\right] f^{(0)}_\text{LE} \; ,
    \label{ainfty}
\end{equation}
a relation which is independent of pseudo gauge. Comparing \eq\eqref{ainfty} with \eq\eqref{azub}, we find agreement if we set $\xi=(E_p+m)/E_p$ and $\Omega^{\mu\nu}=\varpi^{\mu\nu}$ in \eq\eqref{ainfty}. 

However, as noted above,  the polarization in the Zubarev approach depends on the choice of pseudo gauge~\cite{Buzzegoli:2021wlg}, which a priori spoils the physical interpretation of the result.  Let us recall that, in the relativistic case, the splitting of the total angular-momentum tensor $J^{\lambda,\mu\nu}$ into orbital part and spin part,
\begin{equation}
    J^{\lambda,\mu\nu}=x^\mu T^{\lambda\nu}-x^\nu T^{\lambda\mu}+\hbar\, S^{\lambda,\mu\nu}
\end{equation}
is ambiguous, resulting in the so-called pseudo-gauge freedom in the definition of the energy-momentum tensor $T^{\mu\nu}$ and the spin tensor $S^{\lambda,\mu\nu}$~\cite{Hehl:1976vr}. 
Different sets of tensors are related through the transformations
\begin{align}
     T^{\prime\mu\nu}&= T^{\mu\nu}+\frac{\hbar}{2}\partial_{\lambda}(\Phi^{\lambda\mu\nu}+\Phi^{\nu\mu\lambda}+\Phi^{\mu\nu\lambda})\; , \\
  S^{\prime\lambda,\mu\nu}&= S^{\lambda,\mu\nu}-\Phi^{\lambda\mu\nu}\; ,
\end{align}
where $\Phi^{\lambda\mu\nu}$ is the pseudo-gauge potential. Note that the pseudo-gauge ambiguity concerns the spin tensor, whereas the spin vector \eqref{amu}, on which the formalism of the present paper is based, is defined \textit{independently} of the pseudo gauge choice. In the Zubarev approach, the density matrix is split into a local-equilibrium and a dissipative part. Results for physical quantities are derived only from the local-equilibrium density matrix, which is an explicit function of the energy-momentum and spin tensors and hence is a pseudo-gauge dependent quantity~\cite{Becattini:2018duy,Speranza:2020ilk}, while the dissipative part is neglected. Thus, results for physical quantities obtained in this way depend in general on the pseudo gauge. On the other hand, kinetic theory is derived without referring to an energy-momentum or spin tensor, and it is therefore pseudo-gauge independent. The comparison between the two approaches, which both are an approximation for the same underlying exact theory, can therefore provide an interpretation of the pseudo-gauge dependent terms in the Zubarev approach. 

Equation \eqref{azub} corresponds to the polarization derived using the Belinfante pseudo gauge, where the spin tensor is zero and no spin potential is coupled. If one instead uses the canonical set of tensors, whose form is directly derived from Noether's theorem, one obtains~\cite{Buzzegoli:2021wlg}
\begin{align}
\A^\mu_{Z,C} &= -\frac{\hbar}{4m} \left[ \tilde{\varpi}^{\mu\nu}p_\nu+\frac{1}{E_p} \epsilon^{\mu\lambda\rho\sigma}p_\rho t_\sigma p^\nu\partial_{(\lambda} \beta_{\nu)}-\epsilon^{\mu\rho\sigma\tau}p_\tau \left(\varpi_{\rho\sigma}-\Omega_{\rho\sigma}\right)+\frac{2}{E_p}\epsilon^{\mu\lambda\sigma\tau}p^\rho p_\tau \left(\varpi_{\rho\sigma}-\Omega_{\rho\sigma}\right)t_\lambda\right]  f^{(0)}_\text{LE} \n\\
&=  -\frac{\hbar}{4m}\left[\tilde{\Omega}^{\mu\nu}p_\nu+\frac{2}{E_p} \epsilon^{\mu\lambda\rho\sigma}p_\rho t_\sigma p^\nu(\partial_{\lambda} \beta_{\nu}+\Omega_{\lambda\nu}) \right] f^{(0)}_\text{LE}\;.
\end{align}
Comparing with the asymptotic polarization \eqref{ainfty}, the two expression agree for $\xi=(E_p+m)/E_p$ and this time for any spin potential.

Going one step further, we may also compare to the results of Ref.~\cite{Buzzegoli:2021wlg} for a pseudo-gauge transformation with
\begin{equation}
\Phi_{\zeta}^{\lambda\mu\nu}= {\zeta}\, \Phi^{\lambda\mu\nu}_\text{HW}
\label{phigamma}
\end{equation}
with $\zeta$ being a number with $0\leq\zeta\leq1$. Here we introduced the pseudo-gauge potential for the Hilgevoord-Wouthuysen (HW) pseudo-gauge transformation~\cite{HILGEVOORD19631}, see Ref.~\cite{Buzzegoli:2021wlg} for the explicit form and, e.g., Ref.~\cite{Speranza:2020ilk} for a detailed discussion. In the HW pseudo-gauge, the antisymmetric part of the energy-momentum tensor vanishes for free fields and the spin tensor is conserved separately. Using a pseudo-gauge transformation with $\Phi^{\lambda\mu\nu}$ given by \eq\eqref{phigamma}, the polarization obtained from the Zubarev formalism reads~\cite{Buzzegoli:2021wlg} 
\begin{align}
\A^\mu_{Z,\zeta}&=\A^\mu_{Z,C}+\zeta \frac{\hbar}{2mE_p} \epsilon^{\mu\lambda\rho\sigma}p_\rho  t_\sigma p^\nu (\partial_\lambda\beta_\nu+\Omega_{\lambda\nu})\n\\
&= -\frac{\hbar}{4m}\left[\tilde{\Omega}^{\mu\nu}p_\nu+(1-\zeta)\frac{2}{E_p} \epsilon^{\mu\lambda\rho\sigma}p_\rho t_\sigma p^\nu(\partial_{\lambda} \beta_{\nu}+\Omega_{\lambda\nu}) \right] f^{(0)}_\text{LE} \; .
\end{align}
We see from comparison to \eq\eqref{ainfty} that $\zeta$ then is directly related to $\xi=(1-\zeta)(E_p+m)/E_p$.
In particular, the long-time polarization \eqref{ainfty} corresponds to the polarization obtained from Zubarev with the HW pseudo-gauge  if $\xi=0$. \bbb{Recalling that the polarization obtained from the Zubarev formalism is pseudo-gauge dependent, since an explicit form of the spin tensor is used, we conclude that} the HW pseudo-gauge in the Zubarev formalism corresponds to a situation in which the conversion between orbital angular momentum and spin happens very slowly, such that the nonlocal part of the collision term can be neglected and the spin tensor is considered to be conserved. Applying pseudo-gauge transformations with different values of $\zeta$ in \eq\eqref{phigamma}, one can thus choose the equivalent of different time scales for angular-momentum conversion in the Zubarev formalism. We remark that this relation between the pseudo gauge and the angular-momentum conversion time scale is specific to the Zubarev formalism, where only the local-equilibrium part of the density matrix is considered and interaction terms are neglected in general.

\section{Example: rotating cylinder}
\label{cylsec}

\subsection{Nonrelativistic spin kinetic theory and hydrodynamics}
\label{nonrelsec}

In this section, we will demonstrate how the nonlocal collision term \eqref{nlrta} converts orbital angular into polarization by studying the example of a nonrelativistic rotating cylinder. To this end, we first review basic concepts and notations of nonrelativistic spin hydrodynamics. While relativistic spin hydrodynamics is a relatively new development, nonrelativistic hydrodynamics with spin has a long history, and is widely applied, e.g., in the context of micropolar fluids~\cite{Lukaszewicz1999}, spintronics~\cite{takahashi2016spin}, and chiral active fluids~\cite{banerjee2017odd}. Following the conventions of Ref.~\cite{Lukaszewicz1999}, the hydrodynamic equation of motion for the internal angular momentum $\boldsymbol{\ell}$ reads\footnote{We changed the sign of the stress tensor compared to Ref.~\cite{Lukaszewicz1999} in order to be consistent with our definition of the energy-momentum tensor in the nonrelativistic limit.}
\begin{equation}
 \rho (\partial_t + u^j\partial^j )\ell^i=\partial^j C^{ji} + \epsilon^{ijk} T^{kj} \; , \label{eomellgen}
\end{equation}
where $\rho$ is the mass density, $T^{ij}$ is the stress tensor, and $C^{ij}$ is the so-called couple stress tensor. It has been shown in Ref.~\cite{Weickgenannt:2020aaf} that \eq\eqref{eomellgen} can be reproduced as the nonrelativistic limit\footnote{Note that the derivation in \cite{Weickgenannt:2020aaf} uses the so-called Hilgevoord-Wouthuysen (HW) currents~\cite{HILGEVOORD19631} which lead to the correct non-relativistic limit. } of the equation of motion for the spin tensor by making use of the following identifications,\footnote{Note that the factors of $\hbar$ which are not accompanied by a gradient are related to spin and not to the expansion \eqref{hbarexp}. They therefore do not contribute to the $\hbar$ counting.}  
\begin{align}
\rho&\equiv m \left\langle 1\right\rangle\; ,\n\\ 
 \rho \ell^i&\equiv m \frac\hbar2 \left\langle \ms^i \right\rangle\; ,\n\\
C^{ji}&\equiv -\frac\hbar2 \langle \ms^i p^j \rangle+m \frac\hbar2 \langle \ms^i \rangle u^j\; ,\n\\
T^{[ji]}&\equiv m \epsilon^{ijk} \partial_t\left\langle\frac\hbar2 \ms^k\right\rangle + m\epsilon^{ijk}\partial^\ell \left\langle v^\ell \frac\hbar2 \ms^k\right\rangle \; ,
\label{idents}
\end{align}
where we defined $\mathbf{v}\equiv\mathbf{p}/m$ and $\langle\ldots\rangle$ denotes an expectation value with respect to the distribution $f$: $\langle\ldots\rangle\equiv [1/(2\pi\hbar)^3](m^2/2\pi\sqrt{3})\int d^3v\, d^3\ms\, \delta( \boldsymbol{\ms}^2-3)\ldots\, f$. 
We also introduce at zeroth order in $\hbar$ the mass density
\begin{align}
    \rho_0&\equiv m \left\langle 1\right\rangle_0\; ,
    \label{rho}
\end{align}
and the stress tensor
\begin{align}
    T^{ij}_0&\equiv m \left\langle\, v^i v^j \right\rangle_0\; ,
    \label{t0}
\end{align}
where  $\langle\ldots\rangle_0$ is $\langle\ldots\rangle$ for  $f=f^{(0)}$. We define similarly the local-equilibrium quantities with $\langle\ldots\rangle_\text{LE}$ for $f=f_\text{LE}$.

\subsection{Nonrelativistic limit of the NLRTA}

Using that, for $E_p\simeq m$, \eq\eqref{delta} becomes
\begin{equation}
\Delta^i=-\frac{\hbar}{4m^2}\epsilon^{ijk}p^j \ms^k \; ,
\end{equation}
one rewrites the nonrelativistic limit of the Boltzmann equation within the NLRTA \eqref{nlrta} as
\begin{equation}
 \left( m  \partial_t + \p\cdot \ppartial\right)f=-m \frac{f-f_\text{LE}}{\tau_R}+\frac{\hbar\xi}{4m\tau_R} \epsilon^{ijk} p^j p^\ell \ms^i \left(\partial^k \beta u^\ell-\Omega^{k\ell}\right)f^{(0)} \,,
 \label{nlrtanr0}
\end{equation}
 where in $\partial^k \beta u^\ell$, the partial derivative is meant to act on both $\beta$ and $u^\ell$.
Here $f_\text{LE}$ is the local-equilibrium distribution function for a nonrelativistic rotating system. It is  given by
\begin{equation}
    f_\text{LE}\equiv  e^{-\beta E_p-\beta \mathbf{u}\cdot \p} \left(1+\frac\hbar2 \boldsymbol{\Omega}\cdot \boldsymbol{\ms}\right)\; ,
    \label{flenr}
\end{equation}
where $\Omega^{k\ell}\equiv \epsilon^{k\ell m} \Omega^m$. Dividing \eq\eqref{nlrtanr0} by $m$ we get
\begin{equation}
 \left( \partial_t + \vb\cdot \ppartial\right)f=- \frac{f-f_\text{LE}}{\tau_R}+\frac{\hbar\xi}{4\tau_R} \epsilon^{ijk}  v^j v^\ell  \ms^i \left(\partial^k \beta u^\ell- \Omega^{k\ell}\right)f^{(0)}\; .
 \label{nlrtanr}
\end{equation}

We may obtain a compact expression for the antisymmetric part of the stress tensor by inserting \eq\eqref{nlrtanr} into the last identity in \eqs\eqref{idents}. We get  
\begin{align}
    T^{[ji]}&=m \frac\hbar2\epsilon^{ijk} \left(\partial_t\left\langle \ms^k\right\rangle + \partial^\ell \left\langle v^\ell \ms^k\right\rangle \right) \n\\
    &=-\frac{\hbar}{2}\frac{1}{\tau_R}\epsilon^{ijk}m \left[\left\langle \ms^k\right\rangle-\left\langle \ms^k\right\rangle_\text{LE}    -\frac{\hbar\xi}{4} \epsilon^{mn r} \left\langle v^n v^\ell \ms^k \ms^m \right\rangle_{0} \left(\partial^r \beta u^\ell-\Omega^{r\ell}\right) \right]\n\\
    &=-\frac{1}{\tau_R} \epsilon^{ijk}\left[ \rho  \ell^k-m\int_\mathbf{v} \int_\ms \ms^k\frac{\hbar^2}{4}\boldsymbol{\Omega}\cdot \boldsymbol{\ms}f^{(0)}_\text{LE} -\frac{\xi\hbar^2}{4}m\int_\mathbf{v} \left(v^k v^\ell-v^n v^n \delta^{k\ell}\right)  \left(\beta \omega^\ell-\Omega^{\ell} \right)f^{(0)}_\text{LE}\right]\n\\
    &=-\frac{1}{\tau_R}\epsilon^{ijk} \left[ \rho  \left(\ell^k-\frac{\hbar^2}{4}\Omega^k\right)+\xi\frac{\hbar^2}{8}\left(T_0^{k\ell}-\delta^{k\ell} T^{nn}_0\right)\left(\Omega^\ell-\beta\omega^\ell\right)\right]\; ,
    \label{astnr}
\end{align}
where we used \eq\eqref{flenr}, assumed the temperature to be constant and defined $\omega^i\equiv \epsilon^{ijk}\partial^j u^k$. We also use the notation $\int_\mathbf{v}\equiv [1/(2\pi\hbar)^3](m^2/2) \int d^3v$ and $\int_\ms\equiv (1/\sqrt{3}\pi) d^3\ms\, \delta(\boldsymbol{\ms}^2-3)$. Equation \eqref{astnr} constitutes the second term on the right-hand side of the equation of motion for $\ell^i$ in \eq\eqref{eomellgen}, which is responsible for the conversion between orbital angular momentum and spin. 
 Note that for a local collision term with $\xi=0$ the antisymmetric part of the stress tensor \eqref{astnr} vanishes with the matching condition $(\hbar^2/4)\Omega^k=\ell^k$, which in that case is needed to ensure the conservation of spin angular momentum. Here the factor $\hbar^2$ emerges due to the presence of a factor $\hbar$ in the definition of $\ell^k$ in \eqs\eqref{idents}, which does not contribute to the $\hbar$-gradient expansion. It ensures the correct dimension of angular momentum, since $\hbar\Omega^k$ is dimensionless.


\subsection{Kinetic-theory solution for a rotating cylinder}

Consider now a nonrelativistic rigidly rotating cylinder with constant vorticity $\omega^z\equiv\partial^{[x}u^{y]}$, $\omega^x=\omega^y=0$. For spinless particles, this situation corresponds to a global-equilibrium state. Furthermore, even for particles with spin, the system would be in global equilibrium for any constant polarization if the collision term was local, since then the fluid vorticity cannot be converted into spin polarization. However, if the nonlocality of the collision term is taken into account, the situation becomes different. Let us assume that the system is initialized with zero polarization, i.e., with $f$ being independent of $\ms^k$, but with nonzero vorticity $\omega^z>0$ and homogeneous everywhere in the cylinder. We should now see that the nonlocal collision term drives the system to a state in which spin polarization and vorticity are aligned, while conserving total angular momentum. We will quantitatively study this effect in the following.

The initial distribution function is given by
\begin{equation}
    f_\text{in}= e^{-\beta (E_p-\mathbf{u}_\bot\cdot \p_\bot)}
    \label{fin}
\end{equation}
with $\beta$ being \bbb{independent of spatial coordinates}, $\partial_z \mathbf{u}_\bot=0$, and $\partial^{(x}u^{y)}=0$. Formally, this distribution function corresponds to a local-equilibrium distribution function, which does not have contributions at first order in $\hbar$ since we choose the initial polarization to be zero. However, in general, it is not identical to the local-equilibrium distribution function \eqref{fle}, which contains a nonzero spin potential even at the initial time. In fact, $f_\text{LE}\neq f_\text{in}$ at any time, and in particular $f(t=0)\equiv f_\text{in}\neq f_\text{LE}(t=0)$. We will come back to the inital conditions later.
We note that the nonlocality enters the collision term only at first order in $\hbar$, hence, the zeroth-order distribution function is not affected and retains its initial equilibrium form. \bbb{This implies that fluid velocity and temperature obtained from the matching conditions are constant with time.} We may therefore replace $f^{(0)}$ by $f^{(0)}_\text{LE}$ in \eq\eqref{nlrtanr},
\begin{equation}
\left(\partial_t+\mathbf{v}\cdot\boldsymbol{\partial} \right)f=-\frac{f-f_\text{LE}}{\tau_R}+\frac{\hbar}{4\tau_R}\xi \epsilon^{ijk} v^j v^\ell \ms^i(\beta\partial^k u^\ell-\Omega^{k\ell}) f^{(0)}_\text{LE} \; .
    \label{bolbol}
\end{equation}
 Inserting the local-equilibrium distribution \eqref{flenr} into \eq\eqref{bolbol}, one rewrites the Boltzmann equation as
\begin{equation}
\left(\partial_t+\mathbf{v}\cdot\partial\right) f
    =-\frac{1}{\tau_R}\left[f-\left(1+\frac{\hbar}{2} \boldsymbol{\Omega}\cdot\boldsymbol{\ms}\right)\, e^{-\beta (E_p-\mathbf{u}_\bot\cdot \p_\bot)}\right]-\frac{\hbar\xi}{4\tau_R}\left[\mathbf{v}^2 \ms^z-v^z\left( \mathbf{v}\cdot \boldsymbol{\ms}\right)\right] \left(\beta\omega^z-\Omega^z\right) e^{-\beta (E_p-\mathbf{u}_\bot\cdot \p_\bot)}\; .
    \label{bolbolbol}
\end{equation}
Due to the presence of $\ms$-dependent terms on the right-hand side of this equation, a polarization will start to emerge from the vorticity $\omega^z$. To solve this equation of motion for the distribution function, we make the ansatz 
\begin{equation}
    f(t,x,y)=e^{-\beta [E_p-\mathbf{u}_\bot(t,x,y)\cdot \p_\bot]}\left\{1+\frac{\hbar}{2} \left[{\Omega}^z(t)+\delta\Omega^z(t)\right]{\ms}^z+ \frac{\hbar}{4}\left[\mathbf{v}^2 \ms^z-v^z\left( \mathbf{v}\cdot \boldsymbol{\ms}\right)\right] \Lambda(t) \right\}\; , \label{ansatzf}
\end{equation}
where we introduced the parameter $\Lambda(t)$ and a nonequilibrium correction to the spin potential $\delta\Omega^z(t)$. The splitting between $\Omega^z$ and $\delta\Omega^z$ is determined by the matching condition for the total angular momentum, see \eq\eqref{ammatch}. Taking the nonrelativistic limit of \eq\eqref{ammatch} and inserting the ansatz \eqref{ansatzf}, we obtain
\begin{align}
0&=    \int_\mathbf{v}\int_\ms \left(\epsilon^{ijk}\ms^k+\frac12 v^{[j} \epsilon^{i]k\ell} \ms^k v^\ell\right) \left\{\frac12 \delta\Omega^z\ms^z+\frac14\left[\mathbf{v}^2 \ms^z-v^z\left( \mathbf{v}\cdot \boldsymbol{\ms}\right)\right]\left[\Lambda+\xi \left(\beta\omega^z-\Omega^z\right)\right]\right\} e^{-\beta (E_p-\mathbf{u}_\bot\cdot \p_\bot)}\n\\
&=\int_\mathbf{v} \left\{ \epsilon^{ijz}\delta\Omega^z+\frac12v^{[j}\epsilon^{i]z\ell}v^\ell\delta\Omega^z+\frac12\left(\epsilon^{ijz} \mathbf{v}^2-\epsilon^{ijk}v^k v^z-\frac12 v^{[j} \epsilon^{i]z\ell} v^\ell\mathbf{v}^2+\frac12 v^{[j} \epsilon^{i]k\ell} v^\ell v^k v^z\right)\left[\Lambda+\xi \left(\beta\omega^z-\Omega^z\right)\right]\right\}\n\\
&\times e^{-\beta (E_p-\mathbf{u}_\bot\cdot \p_\bot)}\; ,
\end{align}
where the last term vanishes because of the antisymmetry of the epsilon tensor. For $i=z$ or $j=z$ the integral is zero,
since the integrand is odd in $v^z$. For $i=x$, $j=y$ we find 
\begin{align}
   &\int_\mathbf{v} \left\{\delta\Omega^z+\frac12 v^{[y}\epsilon^{x]z\ell}v^\ell\delta\Omega^z+\frac12\left(\mathbf{v}^2-v_z^2-\frac12 v^{[y} \epsilon^{x]z\ell} v^\ell\mathbf{v}^2\right)\left[\Lambda+\xi \left(\beta\omega^z-\Omega^z\right)\right]\right\} e^{-\beta (E_p-\mathbf{u}_\bot\cdot \p_\bot)}\n\\
    &= \int_\mathbf{v} \left\{\delta\Omega^z+\frac12 v^{y}\epsilon^{xzy}v^y\delta\Omega^z-\frac12 v^{x}\epsilon^{yzx}v^x\delta\Omega^z +\frac12\left(\mathbf{v}_\bot^2-\frac12 v^{y} \epsilon^{xzy} v^y\mathbf{v}^2+\frac12v^x\epsilon^{yzx}v^x\mathbf{v}^2\right)\left[\Lambda+\xi \left(\beta\omega^z-\Omega^z\right)\right]\right\}\n\\
    &\times e^{-\beta (E_p-\mathbf{u}_\bot\cdot \p_\bot)}\n\\
    &=\frac12 \int_\mathbf{v} (2-v_y^2-v_x^2)e^{-\beta (E_p-\mathbf{u}_\bot\cdot \p_\bot)} \delta\Omega^z+ \frac14\int_\mathbf{v}  \mathbf{v}_\bot^2 \left(2+\mathbf{v}^2\right)  e^{-\beta (E_p-\mathbf{u}_\bot\cdot \p_\bot)} \left[\Lambda+\xi \left(\beta\omega^z-\Omega^z\right)\right]\; .
    \label{...}
\end{align}
Setting the last line of \eq\eqref{...} to zero determines $\delta\Omega^z$ as a function of $\Lambda$ and $\Omega^z$,
\begin{equation}
    \delta\Omega^z(t,x,y)= \gamma(x,y) \left[\Lambda+\xi \left(\beta\omega^z-\Omega^z\right)\right]\; \label{deltaomegaz}
\end{equation}
with
\begin{equation}
    \gamma\equiv -\frac12 \int_\mathbf{v}  \mathbf{v}_\bot^2 \left(2+\mathbf{v}^2\right)  e^{-\beta (E_p-\mathbf{u}_\bot\cdot \p_\bot)} \left[\int_\mathbf{v} \left(2-\mathbf{v}_\bot^2\right)e^{-\beta (E_p-\mathbf{u}_\bot\cdot \p_\bot)}  \right]^{-1} \; .
    \label{gamma}
\end{equation}

In the next step, we derive the equations of motion for $\Omega^z$ and $\Lambda$. Inserting \eq\eqref{ansatzf} into \eq\eqref{bolbolbol}, we obtain
\begin{align}
    &\partial_t \left(e^{-\beta (E_p-\mathbf{u}_\bot\cdot \p_\bot)}\left\{1+\frac{\hbar}{2} \left[{\Omega}^z(t)+\delta\Omega^z(t)\right]{\ms}^z+ \frac\hbar4\left[\mathbf{v}^2 \ms^z-v^z\left( \mathbf{v}\cdot \boldsymbol{\ms}\right)\right] \Lambda(t) \right\}\right)\n\\
    &=\frac{1}{\tau_R}\left\{-\frac\hbar2 \delta\Omega^z-\frac\hbar4\left[\mathbf{v}^2 \ms^z-v^z\left( \mathbf{v}\cdot \boldsymbol{\ms}\right)\right] \Lambda(t)-\frac\hbar4\xi\left[\mathbf{v}^2 \ms^z-v^z\left( \mathbf{v}\cdot \boldsymbol{\ms}\right)\right] \left[\beta\omega^z-\Omega^z(t)\right]\right\} e^{-\beta (E_p-\mathbf{u}_\bot\cdot \p_\bot)}\; .
    \label{partf}
\end{align}
As already anticipated, we find that $\mathbf{u}_\bot$ does not depend on $t$, since it is defined at zeroth order in $\hbar$ and hence not affected by the nonlocality. 
The equations of motion for the time-dependent quantities are obtained from \eq\eqref{partf} as
\begin{align}
    \partial_t [\Omega^z(t)+\delta\Omega^z(t)]&=-\frac{1}{\tau_R}\delta\Omega^z(t)\; ,\n\\
    \partial_t \Lambda(t)&= -\frac{1}{\tau_R}\left[\Lambda+\xi \left(\beta\omega^z-\Omega^z\right)\right]
    \label{eomomlam} \; .
\end{align}
Inserting $\delta\Omega^z$ from \eq\eqref{deltaomegaz} into \eq\eqref{eomomlam}, we find
\begin{align}
   (1-\gamma\xi) \partial_t \Omega^z(t)+\gamma\partial_t\Lambda(t)&=-\frac{1}{\tau_R}\gamma \left[\Lambda+\xi \left(\beta\omega^z-\Omega^z\right)\right]\; ,\n\\
    \partial_t \Lambda(t)&= -\frac{1}{\tau_R}\left[\Lambda+\xi \left(\beta\omega^z-\Omega^z\right)\right]\; .
\end{align}
Finally, inserting the second line into the first line, we obtain
\begin{align}
   (1-\gamma\xi) \partial_t \Omega^z(t)+\gamma\partial_t\Lambda(t)&=\gamma \partial_t \Lambda(t)\; ,\n\\
    \partial_t\Lambda(t)&= -\frac{1}{\tau_R}\left[\Lambda+\xi \left(\beta\omega^z-\Omega^z\right)\right]\; .
\end{align}
It follows that $\Omega^z\equiv\Omega^z_0$ is constant, and with $\Lambda(0)=0$ the solution of the equations of motion reads
\begin{align}
    \Omega^z(t)&=\Omega^z_0\; ,\n\\
    \Lambda(t)&=  -\xi (\beta\omega^z-\Omega^z_0) \left(1-e^{-t/\tau_R} \right)\; , \n\\
    \delta\Omega^z(t)&= \gamma\, \left[-\xi (\beta\omega^z-\Omega_0^z) \left(1-e^{-t/\tau_R} \right)+\xi (\beta\omega^z-\Omega^z_0)\right]=\gamma  \xi (\beta\omega^z-\Omega^z_0) e^{-t/\tau_R}\; .
    \label{solkin}
\end{align}
Imposing that the initial polarization vanishes yields
\begin{equation}
    0=\Omega^z(0)+\delta\Omega^z(0)=\Omega^z_0+\gamma  \xi (\beta\omega^z-\Omega^z_0)= \left(1-\gamma  \xi\right)\Omega^z_0+\gamma \xi \beta \omega^z
\end{equation}
and therefore
\begin{equation}
    \Omega^z_0= -\frac{\gamma\xi}{1-\gamma\xi}\beta\omega^z \; .
    \label{omega0}
\end{equation}
Inserting this into \eqs\eqref{solkin}, we arrive at the solution
\begin{align}
    \Omega^z(t)&=-\frac{\gamma\xi}{1-\gamma\xi}\beta\omega^z\; ,\n\\
    \Lambda(t)&= -\frac{\xi}{1-\gamma\xi}\beta\omega^z \left(1-e^{-t/\tau_R} \right)\; , \n\\
    \delta\Omega^z(t)&=  \frac{\gamma\xi}{1-\gamma\xi}\beta\omega^z e^{-t/\tau_R}\; .
    \label{solkin2}
\end{align}
We conclude that the nonlocality of the collision term is responsible for the development of a polarization/ internal angular momentum $\ell^z$ from the vorticity $\omega^z$. 
Inserting the distribution function \eqref{ansatzf} with the solution \eqref{solkin} into the second identity in \eqs\eqref{idents}, we obtain
\begin{align}
    \ell^z(t)&= -\frac\hbar2 \frac{m}{\rho} \frac{\xi}{1-\gamma\xi} \int_\mathbf{v} \int_{\ms} \ms^z e^{-\beta [E_p-\mathbf{u}_\bot(t,x,y)\cdot \p_\bot]}\left\{\frac{\hbar}{2} \gamma{\ms}^z+\frac\hbar4\left[\mathbf{v}^2 \ms^z-v^z\left( \mathbf{v}\cdot \boldsymbol{\ms}\right)\right]   \right\} \left(1-e^{-t/\tau_R} \right) \beta \omega^z\n\\
    &= \hbar^2 \frac{m}{\rho} \frac{\eta\xi}{1-\gamma\xi} \left(1-e^{-t/\tau_R} \right) \beta \omega^z\; ,
    \label{ellkinsol}
\end{align}
where we defined the thermodynamic integral 
\begin{equation}
    \eta\equiv -\frac14 \int_\mathbf{v} e^{-\beta [E_p-\mathbf{u}_\bot(t,x,y)\cdot \p_\bot]} \left(2\gamma+\mathbf{v}_\bot^2\right)\; .
    \label{iota}
\end{equation}

In this example, the nonlocal collision term quickly converts orbital angular momentum into spin, the polarization saturates exponentially to its final form. The time scale on which this happens is determined by the relaxation time $\tau_R$ for spin diffusion, while the ratio of the nonlocality time scale to $\tau_R$, which is given by $\xi$ and corresponds to the "effectiveness" of converting angular momentum into spin, enters only the final form of the polarization. However, the polarization is not simply proportional to $\xi$, since the latter affects the angular-momentum conversion in both directions, and the back conversion from spin to orbital angular momentum prevents the system from reaching a state with arbitrary large polarization.
The local-equilibrium part of the spin potential remains constant and proportional to the vorticity, while the dissipative part, which initially cancels the local-equilibrium part, decays exponentially, such that the total spin potential converges to the vorticity. Also note that the system does not reach local equilibrium, the distribution function does not approach the from \eqref{flenr} for $t\rightarrow\infty$. Instead, the distribution function converges to an asymptotic form equivalent to \eq\eqref{finfty1}.

For a local collision term, or in other words, in  the limit $\xi\rightarrow0$, where the process of aligning spin vorticity is much slower than the spin diffusion, no polarization develops. On the other hand, in the limit of large $\xi$, the local-equilibrium part of the spin potential \eqref{omega0} is given by
\begin{equation}
    \Omega_0^z\simeq \beta\omega^z
\end{equation}
and the solution of the equations of motion for $\xi\rightarrow\infty$ reads
\begin{align}
    \Omega^z(t)&=\beta\omega^z\; ,\n\\
    \Lambda(t)&= \frac{1}{\gamma} \beta\omega^z \left(1-e^{-t/\tau_R} \right)\; , \n\\
    \delta\Omega^z(t)&= -\beta\omega^z e^{-t/\tau_R}\; .
    \label{solkinfast}
\end{align}
Thus, if the process of converting vorticity into spin happens much faster than spin diffusion, the solution becomes independent of $\xi$, and in particular is finite for $\xi\rightarrow\infty$.

\subsection{Hydrodynamic description of the rotating cylinder}

Although the example of the rotating cylinder cannot be described through an expansion around local equilibrium, we may equivalently derive the solution for $\ell^z$ by studying the hydrodynamic equation of motion \eqref{eomellgen} supplemented with an equation of motion for an additional current appearing in the matching condition, as we will see in the following. The reason for the equivalence of the kinetic and the hydrodynamic approach lies in the highly symmetric setup, in which the dynamics is determined only by a small number of moments of the distribution function. In this subsection, we outline the derivation of the solution \eqref{ellkinsol} for the polarization in the hydrodynamic description, meaning that, in contrast to the previous calculation, we now do not directly solve the Boltzmann equation for the microscopic distribution function, but study equations of motion for macroscopic quantities.

To solve the equation of motion for the internal angular momentum \eqref{eomellgen} for $i=z$, we need to express the quantities of the right-hand side in terms of $\ell^z$. 
The antisymmetric part of the stress tensor $T^{[xy]}$ is, according to \eq\eqref{astnr}, given by
\begin{align}
    T^{[xy]}&= \frac{1}{\tau_R}\left[ \rho  \left(\ell^z-\frac{\hbar^2}{4}\Omega^z\right)+\xi\frac{\hbar^2}{8}\left(T_0^{zz}- T^{nn}_0\right)\left(\Omega^z-\beta\omega^z\right)\right]\; .
  \label{antisymstress}
  \end{align}
We also note that the first term on the right-hand side of \eq\eqref{eomellgen} vanishes when using the definition of the couple stress tensor in \eqs\eqref{idents} and the rotational symmetry of the system. Hence, the evolution of the internal angular momentum in this example is couple-stress free. Inserting \eq\eqref{antisymstress} into \eq\eqref{eomellgen}, we obtain the equation of motion
\begin{align}
    \rho\partial_t \ell^z&= -\frac{1}{\tau_R}\left[ \rho  \left(\ell^z-\frac{\hbar^2}{4}\Omega^z\right)-\xi\frac{\hbar^2}{8}\left(T_0^{xx}+T_0^{yy} \right)\left(\Omega^z-\beta\omega^z\right)\right]\; .
    \label{eomellll}
\end{align}
  
To proceed, we need to determine $\Omega^z$ from the matching condition \eqref{ammatch} for $\mu=x$, $\nu=y$ in the nonrelativistic limit. 
Note that the latter involves, in addition to the stress tensor and the internal angular momentum, the hydrodynamic quantity
\begin{equation}
  \rho  \mathcal{L}^{z}\equiv m \langle \Delta^{[x} p^{y]} \rangle \; ,
\end{equation}
which is related to the orbital angular momentum transfer in a particle collision. Explicitly the matching condition then reads
\begin{align}
   \rho( \mathcal{L}^z+\ell^z)&= \left\langle \Delta^{[x} p^{y]}+\frac\hbar2 \ms^z  \right\rangle_\text{LE}-\xi (\beta\omega^z-\Omega^z)  \left\langle \left(\Delta^{[x} p^{y]}+\frac\hbar2 \ms^z\right) \Delta^z  \right\rangle_0\n\\
    &= \hbar^2 \int_\mathbf{v} \left\{\frac12\Omega^z+\frac14 v^{[y}\epsilon^{x]z\ell}v^\ell\Omega^z-\frac14\left(\mathbf{v}^2-v_z^2+\frac12 v^{[y} \epsilon^{x]z\ell} v^\ell\mathbf{v}^2\right)\xi \left(\beta\omega^z-\Omega^z\right)\right\} e^{-\beta (E_p-\mathbf{u}_\bot\cdot \p_\bot)}\n\\
    &=-\frac14 \hbar^2\int_\mathbf{v} (2-v_y^2-v_x^2)e^{-\beta (E_p-\mathbf{u}_\bot\cdot \p_\bot)} \Omega^z- \frac18\hbar^2\int_\mathbf{v}  \mathbf{v}_\bot^2 \left(2-\mathbf{v}^2\right)  e^{-\beta (E_p-\mathbf{u}_\bot\cdot \p_\bot)} \xi \left(\beta\omega^z-\Omega^z\right)\; .
    \label{matchnrexpl}
\end{align}
The equation of motion for $\mL^z$ is obtained from the Boltzmann equation \eqref{bolbolbol} as
\begin{align}
    \partial_t \rho \mL^z&= -\frac{1}{\tau_R} \rho\mL^z + \frac{1}{\tau_R} m  \left\langle \Delta^{[x} p^{y]} \right\rangle_\text{LE}+\frac{\hbar}{4}\frac{\xi}{\tau_R}m(\beta\omega^z-\Omega^z) \left\langle \left[\mathbf{v}^2 \ms^z-v^z\left( \mathbf{v}\cdot \boldsymbol{\ms}\right)\right] \Delta^{[x} p^{y]} \right\rangle_0 \n\\
    &= -\frac{1}{\tau_R} \rho\mL^z+ \frac{1}{\tau_R} \rho( \mathcal{L}^z+\ell^z)-\frac{\hbar^2}{2}\rho \Omega^z +\frac{\hbar^2}{4} m \xi (\beta\omega^z-\Omega^z) \int_\mathbf{v} \mathbf{v}_\bot^2  e^{-\beta (E_p-\mathbf{u}_\bot\cdot \p_\bot)}\n\\
    &= \frac{1}{\tau_R}\left[\rho\ell^z+\frac{\hbar^2}{4}\rho \Omega^z + \frac{\hbar^2}{8}\xi (\beta\omega^z-\Omega^z) \left(T_0^{xx}+T_0^{yy}\right) \right]\; ,
    \label{eomcall}
\end{align}
where we inserted the matching condition \eqref{matchnrexpl} in the second step and used $f^{(0)}=f^{(0)}_\text{LE}$ with $f_\text{LE}$ given by \eq\eqref{flenr}. We also made use of the definitions of the mass density and the stress tensor in \eqs\eqref{rho} and \eqref{t0}, respectively. Now taking the time derivative of \eq\eqref{matchnrexpl}, using that $\omega^z$ is constant and inserting \eqs\eqref{eomellll} and \eqref{eomcall}, we find
\begin{align}
&\hbar^2\left[-\frac14 \int_\mathbf{v} (2-\mathbf{v}^2_\bot)e^{-\beta (E_p-\mathbf{u}_\bot\cdot \p_\bot)} + \frac18\xi\int_\mathbf{v}  \mathbf{v}_\bot^2 \left(2-\mathbf{v}^2\right)  e^{-\beta (E_p-\mathbf{u}_\bot\cdot \p_\bot)} \right]\partial_t \Omega^z\n\\
&= \partial_t \rho( \mathcal{L}^z+\ell^z)\n\\
&=\frac{1}{\tau_R}\left[\rho\ell^z-\frac{\hbar^2}{4}\rho \Omega^z + \frac{\hbar^2}{8}\xi (\beta\omega^z-\Omega^z) \left(T_0^{xx}+T_0^{yy}\right)\right]  -\frac{1}{\tau_R}\left[ \rho  \left(\ell^z-\frac{\hbar^2}{4}\Omega^z\right)-\frac{\hbar^2}{8}\xi\left(T_0^{xx}+T_0^{yy}\right)\left(\Omega^z-\beta\omega^z\right)\right]\n\\
&=0\; .
\end{align}
Therefore, we may conclude that $\Omega^z\equiv\Omega^z_0$ is constant and  \eq\eqref{eomellll} is solved by
\begin{equation}
    \ell^z(t)= \left[\frac{\hbar^2}{4}\Omega_0^z-\frac{\xi}{\rho}\frac{\hbar^2}{8}(T_0^{xx}+T_0^{yy})(\Omega^z_0-\beta\omega^z)\right]\left(1-e^{-t/\tau_R}\right) \; ,
    \label{ellhydsol}
\end{equation}
where we imposed the initial condition $\ell^z(0)=0$. Note that to uniquely determine $\ell^z$ we need to specify $\Omega_0^z$, i.e., the initial condition for $\Omega^z$. Using in \eq\eqref{ellhydsol} the same initial condition as for the kinetic-theory solution given by \eq\eqref{omega0}, we recover \eq\eqref{ellkinsol}.~\footnote{The same initial condition could be obtained by solving the equation of motion \eqref{eomcall} for $\mathcal{L}^z$ and imposing $\mathcal{L}^z(0)=0$.}
Therefore, the exact solution for the internal angular momentum may be derived from spin hydrodynamics equivalently to kinetic theory in this special case. The reason for this equivalence lies in the strong symmetry and the limited number of degrees of freedom of the system. In contrast to general dissipative hydrodynamics, the exact system of equations of motion for the relevant macroscopic quantities $\ell^z$ and $\mathcal{L}^z$ is closed already, and thus a truncation is not needed. We remark that the equation of motion \eqref{eomellll} for the internal angular momentum constitutes the conservation of total angular momentum of the fluid. Hence, the increase of polarization is automatically accompanied by a decrease of orbital angular momentum. Note that the latter is contained not only in the vorticity $\omega^z$, which is defined at zeroth order in $\hbar$ and thus constant, but also in the stress tensor at second order, which changes with time.

We may also ask how many particles will be polarized in the final state or equivalently how large the average polarization per particle will be. To obtain the fraction $N_p$ of polarized particles to the total particle number, we only need to divide the internal angular momentum \eqref{ellkinsol} by the polarization of one spin-1/2 particle, namely $\hbar/2$. In the following, we will estimate the order of magnitude of $N_p$. For simplicity, we consider the limit $\xi\rightarrow\infty$. We obtain from \eq\eqref{ellkinsol} for $t/\tau_R\rightarrow\infty$
\begin{align}
N_p&=    -2\hbar \frac{m}{\rho} \frac{\eta}{\gamma} \beta \omega^z  \n\\
&= \hbar  \beta \omega^z \left\{1-\frac{m}{\rho} \frac14 \left[\int_\mathbf{v}  \mathbf{v}_\bot^2 \left(2+\mathbf{v}^2\right)  e^{-\beta (E_p-\mathbf{u}_\bot\cdot \p_\bot)}\right]^{-1} \left[\int_\mathbf{v} \left(2-\mathbf{v}_\bot^2\right)e^{-\beta (E_p-\mathbf{u}_\bot\cdot \p_\bot)}  \right] \int_\mathbf{v} e^{-\beta [E_p-\mathbf{u}_\bot(t,x,y)\cdot \p_\bot]} \mathbf{v}_\bot^2   \right\}\n\\
&\simeq \hbar  \beta \omega^z \left[1-\frac{m}{\rho} \frac18 \int_\mathbf{v} \left(2-\mathbf{v}_\bot^2\right)e^{-\beta (E_p-\mathbf{u}_\bot\cdot \p_\bot)}  \right]  \n\\
&\simeq \frac34\hbar\beta\omega^z\; ,
\label{estimate}
\end{align}
where we inserted \eqs\eqref{iota} and \eqref{gamma} for $\eta$ and $\gamma$, respectively. As an estimate for a nonrelativistic system with typical $v\ll1$, we neglected terms $\sim v^4$ in the first approximation and terms $\sim v^2$ in the second approximation. Consider for example superfluid \ce{^3}He at a temperature of $1/\beta\sim 10^{-8}$ eV in a cryostat which rotates with $\hbar\omega\sim \hbar\times 1 s^{-1}\sim 10^{-15}$ eV, see e.g.\ Ref.~\cite{vollhardt2013superfluid}, then $N_p\sim 10^{-7}$, which means that only a very small fraction of particles is polarized.  The effect can be increased by lowering the temperature or increasing the rotation speed. The latter is bound due to the restriction to nonrelativistic systems. On the other hand, the temperature can in principle be arbitrarily low. However, we recall that we neglected Fermi-Dirac statistics in the distribution function \eqref{flenr}. Describing a system with temperature lower than the particle mass requires to use the Fermi-Dirac distribution function
\begin{equation}
    f_\text{LE}=\frac{1}{1+e^{-\beta\cdot p}}-\frac{e^{-\beta\cdot p}}{(1+e^{-\beta\cdot p})^2}\frac{\hbar}{4}\Omega_{\mu\nu}\Sigma_\ms^{\mu\nu} \; ,
    \label{ffd}
\end{equation}
where the second term emerges from the expansion at first order in $\hbar$. Using this distribution function would not change the results of this section. However, the values of the integrals in \eq\eqref{estimate} would be modified by the following additional factor
\begin{equation}
    \frac{m}{\rho} \int_\mathbf{v} \frac{e^{-\beta\cdot p}}{(1+e^{-\beta\cdot p})^2} \; .
\end{equation}
This additional factor will damp $N_p$ for small $\beta$ and prevent the whole fraction from getting larger than one.

\section{Conclusions}
\label{concsec}

In this paper, we provided a simple approximation for the nonlocal collision term in spin kinetic theory, called the NLRTA. While being identical to the well-known RTA at zeroth order in a power expansion in $\hbar$, the NLRTA features a new term at first order in $\hbar$, which vanishes for the local-equilibrium distribution function \eqref{fle} only if the global-equilibrium conditions \eqref{globeq} are fulfilled, in agreement with the microscopic nonlocal collision term. We then performed a fast-relaxation expansion in powers of the relaxation time multiplied by a gradient. At leading order in this expansion, the system is described by the asymptotic distribution function, instead of the local-equilibrium distribution function. Higher orders in this expansion may be calculated systematically analogously to a Chapman-Enskog expansion. 

Comparing the polarization obtained from the asymptotic distribution function to results from the Zubarev formalism, we found agreement if the ratio $\xi$ of the time scale associated with angular-momentum conversion to the time scale of spin diffusion in the NLRTA is identified with a parameter related to pseudo-gauge choice in the Zubarev approach. This provides an interpretation of the pseudo-gauge dependent terms in the Lambda polarization recently computed in the context of heavy-ion collisions. Adopting the viewpoint that a physical meaning can be assigned to a pseudo-gauge choice in a specific context, and taking into account that the ``local-equilibrium" polarization obtained from the Zubarev formalism is only an approximation for the ``true" polarization, we may conclude that a suitable choice of the pseudo-gauge  may be suggested by the physical situation. If, e.g., for the considered system the conversion between orbital angular momentum and spin happens much slower than diffusive processes, the polarization obtained from the Zubarev approach using the HW pseudo gauge will provide the best approximation for the true polarization of this system. On the other hand, if both processes happen on the same time scale, the canonical pseudo gauge appears most suitable. Thus, a certain pseudo-gauge choice in the Zubarev approach corresponds to a certain effectiveness of the conversion of orbital angular momentum into spin and vice versa.

We also studied the example of a nonrelativistic rigidly rotating cylinder. Using the NLRTA, we were able to explicitly obtain the antisymmetric part of the stress tensor, which drives the conversion between internal and orbital angular momentum and vanishes if the internal angular momentum is proportional to the spin potential and the spin potential to the vorticity. We then exactly solved the Boltzmann equation with the NLRTA for this example with zero initial polarization. The result shows that the nonlocal part of the collision term induces spin polarization from the fluid vorticity. The spin potential can be written as the sum of a local-equilibrium and a dissipative part, where the former is constant and proportional to the vorticity, while the latter initially cancels the local-equilibrium part, but decays exponentially with time. Thus, the internal angular momentum exponentially approaches its final form, which is also  proportional to the vorticity with the proportionality coefficient depending on $\xi$: the larger $\xi$, the more effective the conversion between orbital and internal angular momentum compared to the competing process of diffusion, and the larger the final polarization. However, the polarization is bound from above for large $\xi$, with the upper limit determined by the vorticity. In contrast, the relaxation time $\tau_R$ sets the time scale on which the polarization converges to its final form. It is interesting to note that the example discussed here can be exactly described by hydrodynamic equations of motion for micropolar fluids, reproducing the kinetic-theory solution for the internal angular momentum without knowledge of the distribution function. Due to the symmetry of the setup, the number of degrees of freedom of the system is highly limited, and the hydrodynamic equations of motion are closed without truncation. This is in contrast to general hydrodynamic equations, which usually have to be closed by imposing approximate identities between different quantities, such that their solutions cannot be obtained exactly. Due to the established relation to micropolar fluids, it is promising to find more applications of the NLRTA for nonrelativistic fluids with spin.

To summarize, the NLRTA features a much simpler structure than the microscopic nonlocal collision term, while preserving its important properties and physical interpretation. In particular, the asymptotic polarization obtained from the NLRTA reproduces known results from the Zubarev formalism for certain values of $\xi$. Furthermore, the NLRTA allows to exactly solve the Boltzmann equation for the example of a nonrelativistic rotating cylinder and presumably will simplify calculations for spin kinetic theory and hydrodynamics in many situations. It would be interesting to apply it to more examples both in the relativistic and nonrelativistic context.

\section*{Acknowledgements}

N.W.\ acknowledges support by the German National Academy of Sciences Leopoldina through the Leopoldina fellowship program with funding code LPDS 2022-11.

\bibliography{biblio_paper_long}{}

\begin{thebibliography}{83}
\expandafter\ifx\csname natexlab\endcsname\relax\def\natexlab#1{#1}\fi
\expandafter\ifx\csname bibnamefont\endcsname\relax
  \def\bibnamefont#1{#1}\fi
\expandafter\ifx\csname bibfnamefont\endcsname\relax
  \def\bibfnamefont#1{#1}\fi
\expandafter\ifx\csname citenamefont\endcsname\relax
  \def\citenamefont#1{#1}\fi
\expandafter\ifx\csname url\endcsname\relax
  \def\url#1{\texttt{#1}}\fi
\expandafter\ifx\csname urlprefix\endcsname\relax\def\urlprefix{URL }\fi
\providecommand{\bibinfo}[2]{#2}
\providecommand{\eprint}[2][]{\url{#2}}

\bibitem[{\citenamefont{Barnett}(1935)}]{Barnett:1935}
\bibinfo{author}{\bibfnamefont{S.~J.} \bibnamefont{Barnett}},
  \bibinfo{journal}{Rev. Mod. Phys.} \textbf{\bibinfo{volume}{7}},
  \bibinfo{pages}{129} (\bibinfo{year}{1935}).

\bibitem[{\citenamefont{Lukaszewicz}(1999)}]{Lukaszewicz1999}
\bibinfo{author}{\bibfnamefont{G.}~\bibnamefont{Lukaszewicz}},
  \emph{\bibinfo{title}{Micropolar Fluids, Theory and Applications}}
  (\bibinfo{publisher}{Birkh\"auser Boston}, \bibinfo{year}{1999}).

\bibitem[{\citenamefont{Banerjee et~al.}(2017)\citenamefont{Banerjee, Souslov,
  Abanov, and Vitelli}}]{banerjee2017odd}
\bibinfo{author}{\bibfnamefont{D.}~\bibnamefont{Banerjee}},
  \bibinfo{author}{\bibfnamefont{A.}~\bibnamefont{Souslov}},
  \bibinfo{author}{\bibfnamefont{A.~G.} \bibnamefont{Abanov}},
  \bibnamefont{and} \bibinfo{author}{\bibfnamefont{V.}~\bibnamefont{Vitelli}},
  \bibinfo{journal}{Nature communications} \textbf{\bibinfo{volume}{8}},
  \bibinfo{pages}{1} (\bibinfo{year}{2017}).

\bibitem[{\citenamefont{Takahashi et~al.}(2016)\citenamefont{Takahashi, Matsuo,
  Ono, Harii, Chudo, Okayasu, Ieda, Takahashi, Maekawa, and
  Saitoh}}]{takahashi2016spin}
\bibinfo{author}{\bibfnamefont{R.}~\bibnamefont{Takahashi}},
  \bibinfo{author}{\bibfnamefont{M.}~\bibnamefont{Matsuo}},
  \bibinfo{author}{\bibfnamefont{M.}~\bibnamefont{Ono}},
  \bibinfo{author}{\bibfnamefont{K.}~\bibnamefont{Harii}},
  \bibinfo{author}{\bibfnamefont{H.}~\bibnamefont{Chudo}},
  \bibinfo{author}{\bibfnamefont{S.}~\bibnamefont{Okayasu}},
  \bibinfo{author}{\bibfnamefont{J.}~\bibnamefont{Ieda}},
  \bibinfo{author}{\bibfnamefont{S.}~\bibnamefont{Takahashi}},
  \bibinfo{author}{\bibfnamefont{S.}~\bibnamefont{Maekawa}}, \bibnamefont{and}
  \bibinfo{author}{\bibfnamefont{E.}~\bibnamefont{Saitoh}},
  \bibinfo{journal}{Nature Physics} \textbf{\bibinfo{volume}{12}},
  \bibinfo{pages}{52} (\bibinfo{year}{2016}).

\bibitem[{\citenamefont{Liang and Wang}(2005)}]{Liang:2004ph}
\bibinfo{author}{\bibfnamefont{Z.-T.} \bibnamefont{Liang}} \bibnamefont{and}
  \bibinfo{author}{\bibfnamefont{X.-N.} \bibnamefont{Wang}},
  \bibinfo{journal}{Phys. Rev. Lett.} \textbf{\bibinfo{volume}{94}},
  \bibinfo{pages}{102301} (\bibinfo{year}{2005}), \bibinfo{note}{[Erratum:
  Phys. Rev. Lett.96,039901(2006)]}, \eprint{nucl-th/0410079}.

\bibitem[{\citenamefont{Voloshin}(2004)}]{Voloshin:2004ha}
\bibinfo{author}{\bibfnamefont{S.~A.} \bibnamefont{Voloshin}}
  (\bibinfo{year}{2004}), \eprint{nucl-th/0410089}.

\bibitem[{\citenamefont{Betz et~al.}(2007)\citenamefont{Betz, Gyulassy, and
  Torrieri}}]{Betz:2007kg}
\bibinfo{author}{\bibfnamefont{B.}~\bibnamefont{Betz}},
  \bibinfo{author}{\bibfnamefont{M.}~\bibnamefont{Gyulassy}}, \bibnamefont{and}
  \bibinfo{author}{\bibfnamefont{G.}~\bibnamefont{Torrieri}},
  \bibinfo{journal}{Phys. Rev. C} \textbf{\bibinfo{volume}{76}},
  \bibinfo{pages}{044901} (\bibinfo{year}{2007}), \eprint{0708.0035}.

\bibitem[{\citenamefont{Becattini et~al.}(2008)\citenamefont{Becattini,
  Piccinini, and Rizzo}}]{Becattini:2007sr}
\bibinfo{author}{\bibfnamefont{F.}~\bibnamefont{Becattini}},
  \bibinfo{author}{\bibfnamefont{F.}~\bibnamefont{Piccinini}},
  \bibnamefont{and} \bibinfo{author}{\bibfnamefont{J.}~\bibnamefont{Rizzo}},
  \bibinfo{journal}{Phys. Rev.} \textbf{\bibinfo{volume}{C77}},
  \bibinfo{pages}{024906} (\bibinfo{year}{2008}), \eprint{0711.1253}.

\bibitem[{\citenamefont{Becattini and Karpenko}(2018)}]{Becattini:2017gcx}
\bibinfo{author}{\bibfnamefont{F.}~\bibnamefont{Becattini}} \bibnamefont{and}
  \bibinfo{author}{\bibfnamefont{I.}~\bibnamefont{Karpenko}},
  \bibinfo{journal}{Phys. Rev. Lett.} \textbf{\bibinfo{volume}{120}},
  \bibinfo{pages}{012302} (\bibinfo{year}{2018}), \eprint{1707.07984}.

\bibitem[{\citenamefont{Adam et~al.}(2018)}]{Adam:2018ivw}
\bibinfo{author}{\bibfnamefont{J.}~\bibnamefont{Adam}} \bibnamefont{et~al.}
  (\bibinfo{collaboration}{STAR}), \bibinfo{journal}{Phys. Rev.}
  \textbf{\bibinfo{volume}{C98}}, \bibinfo{pages}{014910}
  (\bibinfo{year}{2018}), \eprint{1805.04400}.

\bibitem[{\citenamefont{Acharya et~al.}(2020)}]{ALICE:2019aid}
\bibinfo{author}{\bibfnamefont{S.}~\bibnamefont{Acharya}} \bibnamefont{et~al.}
  (\bibinfo{collaboration}{ALICE}), \bibinfo{journal}{Phys. Rev. Lett.}
  \textbf{\bibinfo{volume}{125}}, \bibinfo{pages}{012301}
  (\bibinfo{year}{2020}), \eprint{1910.14408}.

\bibitem[{\citenamefont{Adam et~al.}(2019)}]{STAR:2019erd}
\bibinfo{author}{\bibfnamefont{J.}~\bibnamefont{Adam}} \bibnamefont{et~al.}
  (\bibinfo{collaboration}{STAR}), \bibinfo{journal}{Phys. Rev. Lett.}
  \textbf{\bibinfo{volume}{123}}, \bibinfo{pages}{132301}
  (\bibinfo{year}{2019}), \eprint{1905.11917}.

\bibitem[{\citenamefont{Florkowski
  et~al.}(2019{\natexlab{a}})\citenamefont{Florkowski, Kumar, Ryblewski, and
  Singh}}]{Florkowski:2019qdp}
\bibinfo{author}{\bibfnamefont{W.}~\bibnamefont{Florkowski}},
  \bibinfo{author}{\bibfnamefont{A.}~\bibnamefont{Kumar}},
  \bibinfo{author}{\bibfnamefont{R.}~\bibnamefont{Ryblewski}},
  \bibnamefont{and} \bibinfo{author}{\bibfnamefont{R.}~\bibnamefont{Singh}},
  \bibinfo{journal}{Phys. Rev.} \textbf{\bibinfo{volume}{C99}},
  \bibinfo{pages}{044910} (\bibinfo{year}{2019}{\natexlab{a}}),
  \eprint{1901.09655}.

\bibitem[{\citenamefont{Florkowski
  et~al.}(2019{\natexlab{b}})\citenamefont{Florkowski, Kumar, Ryblewski, and
  Mazeliauskas}}]{Florkowski:2019voj}
\bibinfo{author}{\bibfnamefont{W.}~\bibnamefont{Florkowski}},
  \bibinfo{author}{\bibfnamefont{A.}~\bibnamefont{Kumar}},
  \bibinfo{author}{\bibfnamefont{R.}~\bibnamefont{Ryblewski}},
  \bibnamefont{and}
  \bibinfo{author}{\bibfnamefont{A.}~\bibnamefont{Mazeliauskas}},
  \bibinfo{journal}{Phys. Rev.} \textbf{\bibinfo{volume}{C100}},
  \bibinfo{pages}{054907} (\bibinfo{year}{2019}{\natexlab{b}}),
  \eprint{1904.00002}.

\bibitem[{\citenamefont{Becattini and Lisa}(2020)}]{Becattini:2020ngo}
\bibinfo{author}{\bibfnamefont{F.}~\bibnamefont{Becattini}} \bibnamefont{and}
  \bibinfo{author}{\bibfnamefont{M.~A.} \bibnamefont{Lisa}},
  \bibinfo{journal}{Ann. Rev. Nucl. Part. Sci.} \textbf{\bibinfo{volume}{70}},
  \bibinfo{pages}{395} (\bibinfo{year}{2020}), \eprint{2003.03640}.

\bibitem[{\citenamefont{Banerjee et~al.}(2024)\citenamefont{Banerjee, Bhadury,
  Florkowski, Jaiswal, and Ryblewski}}]{Banerjee:2024xnd}
\bibinfo{author}{\bibfnamefont{S.}~\bibnamefont{Banerjee}},
  \bibinfo{author}{\bibfnamefont{S.}~\bibnamefont{Bhadury}},
  \bibinfo{author}{\bibfnamefont{W.}~\bibnamefont{Florkowski}},
  \bibinfo{author}{\bibfnamefont{A.}~\bibnamefont{Jaiswal}}, \bibnamefont{and}
  \bibinfo{author}{\bibfnamefont{R.}~\bibnamefont{Ryblewski}}
  (\bibinfo{year}{2024}), \eprint{2405.05089}.

\bibitem[{\citenamefont{Becattini
  et~al.}(2013{\natexlab{a}})\citenamefont{Becattini, Chandra, Del~Zanna, and
  Grossi}}]{Becattini:2013fla}
\bibinfo{author}{\bibfnamefont{F.}~\bibnamefont{Becattini}},
  \bibinfo{author}{\bibfnamefont{V.}~\bibnamefont{Chandra}},
  \bibinfo{author}{\bibfnamefont{L.}~\bibnamefont{Del~Zanna}},
  \bibnamefont{and} \bibinfo{author}{\bibfnamefont{E.}~\bibnamefont{Grossi}},
  \bibinfo{journal}{Annals Phys.} \textbf{\bibinfo{volume}{338}},
  \bibinfo{pages}{32} (\bibinfo{year}{2013}{\natexlab{a}}), \eprint{1303.3431}.

\bibitem[{\citenamefont{Liu and Yin}(2021)}]{Liu:2021uhn}
\bibinfo{author}{\bibfnamefont{S.~Y.~F.} \bibnamefont{Liu}} \bibnamefont{and}
  \bibinfo{author}{\bibfnamefont{Y.}~\bibnamefont{Yin}},
  \bibinfo{journal}{JHEP} \textbf{\bibinfo{volume}{07}}, \bibinfo{pages}{188}
  (\bibinfo{year}{2021}), \eprint{2103.09200}.

\bibitem[{\citenamefont{Fu et~al.}(2021)\citenamefont{Fu, Liu, Pang, Song, and
  Yin}}]{Fu:2021pok}
\bibinfo{author}{\bibfnamefont{B.}~\bibnamefont{Fu}},
  \bibinfo{author}{\bibfnamefont{S.~Y.~F.} \bibnamefont{Liu}},
  \bibinfo{author}{\bibfnamefont{L.}~\bibnamefont{Pang}},
  \bibinfo{author}{\bibfnamefont{H.}~\bibnamefont{Song}}, \bibnamefont{and}
  \bibinfo{author}{\bibfnamefont{Y.}~\bibnamefont{Yin}},
  \bibinfo{journal}{Phys. Rev. Lett.} \textbf{\bibinfo{volume}{127}},
  \bibinfo{pages}{142301} (\bibinfo{year}{2021}), \eprint{2103.10403}.

\bibitem[{\citenamefont{Becattini
  et~al.}(2021{\natexlab{a}})\citenamefont{Becattini, Buzzegoli, and
  Palermo}}]{Becattini:2021suc}
\bibinfo{author}{\bibfnamefont{F.}~\bibnamefont{Becattini}},
  \bibinfo{author}{\bibfnamefont{M.}~\bibnamefont{Buzzegoli}},
  \bibnamefont{and} \bibinfo{author}{\bibfnamefont{A.}~\bibnamefont{Palermo}},
  \bibinfo{journal}{Phys. Lett. B} \textbf{\bibinfo{volume}{820}},
  \bibinfo{pages}{136519} (\bibinfo{year}{2021}{\natexlab{a}}),
  \eprint{2103.10917}.

\bibitem[{\citenamefont{Becattini
  et~al.}(2021{\natexlab{b}})\citenamefont{Becattini, Buzzegoli, Inghirami,
  Karpenko, and Palermo}}]{Becattini:2021iol}
\bibinfo{author}{\bibfnamefont{F.}~\bibnamefont{Becattini}},
  \bibinfo{author}{\bibfnamefont{M.}~\bibnamefont{Buzzegoli}},
  \bibinfo{author}{\bibfnamefont{G.}~\bibnamefont{Inghirami}},
  \bibinfo{author}{\bibfnamefont{I.}~\bibnamefont{Karpenko}}, \bibnamefont{and}
  \bibinfo{author}{\bibfnamefont{A.}~\bibnamefont{Palermo}},
  \bibinfo{journal}{Phys. Rev. Lett.} \textbf{\bibinfo{volume}{127}},
  \bibinfo{pages}{272302} (\bibinfo{year}{2021}{\natexlab{b}}),
  \eprint{2103.14621}.

\bibitem[{\citenamefont{Buzzegoli}(2022)}]{Buzzegoli:2021wlg}
\bibinfo{author}{\bibfnamefont{M.}~\bibnamefont{Buzzegoli}},
  \bibinfo{journal}{Phys. Rev. C} \textbf{\bibinfo{volume}{105}},
  \bibinfo{pages}{044907} (\bibinfo{year}{2022}), \eprint{2109.12084}.

\bibitem[{\citenamefont{Weickgenannt et~al.}(2019)\citenamefont{Weickgenannt,
  Sheng, Speranza, Wang, and Rischke}}]{Weickgenannt:2019dks}
\bibinfo{author}{\bibfnamefont{N.}~\bibnamefont{Weickgenannt}},
  \bibinfo{author}{\bibfnamefont{X.-L.} \bibnamefont{Sheng}},
  \bibinfo{author}{\bibfnamefont{E.}~\bibnamefont{Speranza}},
  \bibinfo{author}{\bibfnamefont{Q.}~\bibnamefont{Wang}}, \bibnamefont{and}
  \bibinfo{author}{\bibfnamefont{D.~H.} \bibnamefont{Rischke}},
  \bibinfo{journal}{Phys. Rev.} \textbf{\bibinfo{volume}{D100}},
  \bibinfo{pages}{056018} (\bibinfo{year}{2019}), \eprint{1902.06513}.

\bibitem[{\citenamefont{Gao and Liang}(2019)}]{Gao:2019znl}
\bibinfo{author}{\bibfnamefont{J.-H.} \bibnamefont{Gao}} \bibnamefont{and}
  \bibinfo{author}{\bibfnamefont{Z.-T.} \bibnamefont{Liang}},
  \bibinfo{journal}{Phys. Rev.} \textbf{\bibinfo{volume}{D100}},
  \bibinfo{pages}{056021} (\bibinfo{year}{2019}), \eprint{1902.06510}.

\bibitem[{\citenamefont{Hattori
  et~al.}(2019{\natexlab{a}})\citenamefont{Hattori, Hidaka, and
  Yang}}]{Hattori:2019ahi}
\bibinfo{author}{\bibfnamefont{K.}~\bibnamefont{Hattori}},
  \bibinfo{author}{\bibfnamefont{Y.}~\bibnamefont{Hidaka}}, \bibnamefont{and}
  \bibinfo{author}{\bibfnamefont{D.-L.} \bibnamefont{Yang}},
  \bibinfo{journal}{Phys. Rev.} \textbf{\bibinfo{volume}{D100}},
  \bibinfo{pages}{096011} (\bibinfo{year}{2019}{\natexlab{a}}),
  \eprint{1903.01653}.

\bibitem[{\citenamefont{Wang et~al.}(2019)\citenamefont{Wang, Guo, Shi, and
  Zhuang}}]{Wang:2019moi}
\bibinfo{author}{\bibfnamefont{Z.}~\bibnamefont{Wang}},
  \bibinfo{author}{\bibfnamefont{X.}~\bibnamefont{Guo}},
  \bibinfo{author}{\bibfnamefont{S.}~\bibnamefont{Shi}}, \bibnamefont{and}
  \bibinfo{author}{\bibfnamefont{P.}~\bibnamefont{Zhuang}},
  \bibinfo{journal}{Phys. Rev. D} \textbf{\bibinfo{volume}{100}},
  \bibinfo{pages}{014015} (\bibinfo{year}{2019}), \eprint{1903.03461}.

\bibitem[{\citenamefont{Weickgenannt
  et~al.}(2021{\natexlab{a}})\citenamefont{Weickgenannt, Speranza, Sheng, Wang,
  and Rischke}}]{Weickgenannt:2020aaf}
\bibinfo{author}{\bibfnamefont{N.}~\bibnamefont{Weickgenannt}},
  \bibinfo{author}{\bibfnamefont{E.}~\bibnamefont{Speranza}},
  \bibinfo{author}{\bibfnamefont{X.-l.} \bibnamefont{Sheng}},
  \bibinfo{author}{\bibfnamefont{Q.}~\bibnamefont{Wang}}, \bibnamefont{and}
  \bibinfo{author}{\bibfnamefont{D.~H.} \bibnamefont{Rischke}},
  \bibinfo{journal}{Phys. Rev. Lett.} \textbf{\bibinfo{volume}{127}},
  \bibinfo{pages}{052301} (\bibinfo{year}{2021}{\natexlab{a}}),
  \eprint{2005.01506}.

\bibitem[{\citenamefont{Liu et~al.}(2020)\citenamefont{Liu, Mameda, and
  Huang}}]{Liu:2020flb}
\bibinfo{author}{\bibfnamefont{Y.-C.} \bibnamefont{Liu}},
  \bibinfo{author}{\bibfnamefont{K.}~\bibnamefont{Mameda}}, \bibnamefont{and}
  \bibinfo{author}{\bibfnamefont{X.-G.} \bibnamefont{Huang}},
  \bibinfo{journal}{Chin. Phys. C} \textbf{\bibinfo{volume}{44}},
  \bibinfo{pages}{094101} (\bibinfo{year}{2020}), \bibinfo{note}{[Erratum:
  Chin.Phys.C 45, 089001 (2021)]}, \eprint{2002.03753}.

\bibitem[{\citenamefont{Weickgenannt
  et~al.}(2021{\natexlab{b}})\citenamefont{Weickgenannt, Speranza, Sheng, Wang,
  and Rischke}}]{Weickgenannt:2021cuo}
\bibinfo{author}{\bibfnamefont{N.}~\bibnamefont{Weickgenannt}},
  \bibinfo{author}{\bibfnamefont{E.}~\bibnamefont{Speranza}},
  \bibinfo{author}{\bibfnamefont{X.-l.} \bibnamefont{Sheng}},
  \bibinfo{author}{\bibfnamefont{Q.}~\bibnamefont{Wang}}, \bibnamefont{and}
  \bibinfo{author}{\bibfnamefont{D.~H.} \bibnamefont{Rischke}},
  \bibinfo{journal}{Phys. Rev. D} \textbf{\bibinfo{volume}{104}},
  \bibinfo{pages}{016022} (\bibinfo{year}{2021}{\natexlab{b}}),
  \eprint{2103.04896}.

\bibitem[{\citenamefont{Sheng et~al.}(2021)\citenamefont{Sheng, Weickgenannt,
  Speranza, Rischke, and Wang}}]{Sheng:2021kfc}
\bibinfo{author}{\bibfnamefont{X.-L.} \bibnamefont{Sheng}},
  \bibinfo{author}{\bibfnamefont{N.}~\bibnamefont{Weickgenannt}},
  \bibinfo{author}{\bibfnamefont{E.}~\bibnamefont{Speranza}},
  \bibinfo{author}{\bibfnamefont{D.~H.} \bibnamefont{Rischke}},
  \bibnamefont{and} \bibinfo{author}{\bibfnamefont{Q.}~\bibnamefont{Wang}},
  \bibinfo{journal}{Phys. Rev. D} \textbf{\bibinfo{volume}{104}},
  \bibinfo{pages}{016029} (\bibinfo{year}{2021}), \eprint{2103.10636}.

\bibitem[{\citenamefont{Wagner et~al.}(2022)\citenamefont{Wagner, Weickgenannt,
  and Rischke}}]{Wagner:2022amr}
\bibinfo{author}{\bibfnamefont{D.}~\bibnamefont{Wagner}},
  \bibinfo{author}{\bibfnamefont{N.}~\bibnamefont{Weickgenannt}},
  \bibnamefont{and} \bibinfo{author}{\bibfnamefont{D.~H.}
  \bibnamefont{Rischke}}, \bibinfo{journal}{Phys. Rev. D}
  \textbf{\bibinfo{volume}{106}}, \bibinfo{pages}{116021}
  (\bibinfo{year}{2022}), \eprint{2210.06187}.

\bibitem[{\citenamefont{Wagner et~al.}(2023)\citenamefont{Wagner, Weickgenannt,
  and Speranza}}]{Wagner:2023cct}
\bibinfo{author}{\bibfnamefont{D.}~\bibnamefont{Wagner}},
  \bibinfo{author}{\bibfnamefont{N.}~\bibnamefont{Weickgenannt}},
  \bibnamefont{and} \bibinfo{author}{\bibfnamefont{E.}~\bibnamefont{Speranza}}
  (\bibinfo{year}{2023}), \eprint{2306.05936}.

\bibitem[{\citenamefont{Fang and Pu}(2024)}]{Fang:2024vds}
\bibinfo{author}{\bibfnamefont{S.}~\bibnamefont{Fang}} \bibnamefont{and}
  \bibinfo{author}{\bibfnamefont{S.}~\bibnamefont{Pu}} (\bibinfo{year}{2024}),
  \eprint{2408.09877}.

\bibitem[{\citenamefont{Wang and Lin}(2024)}]{Wang:2024lis}
\bibinfo{author}{\bibfnamefont{Z.}~\bibnamefont{Wang}} \bibnamefont{and}
  \bibinfo{author}{\bibfnamefont{S.}~\bibnamefont{Lin}} (\bibinfo{year}{2024}),
  \eprint{2411.19550}.

\bibitem[{\citenamefont{Hess and Waldmann}(1966)}]{hess1966kinetic}
\bibinfo{author}{\bibfnamefont{S.}~\bibnamefont{Hess}} \bibnamefont{and}
  \bibinfo{author}{\bibfnamefont{L.}~\bibnamefont{Waldmann}},
  \bibinfo{journal}{Zeitschrift f{\"u}r Naturforschung A}
  \textbf{\bibinfo{volume}{21}}, \bibinfo{pages}{1529} (\bibinfo{year}{1966}).

\bibitem[{\citenamefont{Florkowski
  et~al.}(2018{\natexlab{a}})\citenamefont{Florkowski, Friman, Jaiswal, and
  Speranza}}]{Florkowski:2017ruc}
\bibinfo{author}{\bibfnamefont{W.}~\bibnamefont{Florkowski}},
  \bibinfo{author}{\bibfnamefont{B.}~\bibnamefont{Friman}},
  \bibinfo{author}{\bibfnamefont{A.}~\bibnamefont{Jaiswal}}, \bibnamefont{and}
  \bibinfo{author}{\bibfnamefont{E.}~\bibnamefont{Speranza}},
  \bibinfo{journal}{Phys. Rev.} \textbf{\bibinfo{volume}{C97}},
  \bibinfo{pages}{041901} (\bibinfo{year}{2018}{\natexlab{a}}),
  \eprint{1705.00587}.

\bibitem[{\citenamefont{Florkowski
  et~al.}(2018{\natexlab{b}})\citenamefont{Florkowski, Friman, Jaiswal,
  Ryblewski, and Speranza}}]{Florkowski:2017dyn}
\bibinfo{author}{\bibfnamefont{W.}~\bibnamefont{Florkowski}},
  \bibinfo{author}{\bibfnamefont{B.}~\bibnamefont{Friman}},
  \bibinfo{author}{\bibfnamefont{A.}~\bibnamefont{Jaiswal}},
  \bibinfo{author}{\bibfnamefont{R.}~\bibnamefont{Ryblewski}},
  \bibnamefont{and} \bibinfo{author}{\bibfnamefont{E.}~\bibnamefont{Speranza}},
  \bibinfo{journal}{Phys. Rev.} \textbf{\bibinfo{volume}{D97}},
  \bibinfo{pages}{116017} (\bibinfo{year}{2018}{\natexlab{b}}),
  \eprint{1712.07676}.

\bibitem[{\citenamefont{Florkowski
  et~al.}(2019{\natexlab{c}})\citenamefont{Florkowski, Ryblewski, and
  Kumar}}]{Florkowski:2018fap}
\bibinfo{author}{\bibfnamefont{W.}~\bibnamefont{Florkowski}},
  \bibinfo{author}{\bibfnamefont{R.}~\bibnamefont{Ryblewski}},
  \bibnamefont{and} \bibinfo{author}{\bibfnamefont{A.}~\bibnamefont{Kumar}},
  \bibinfo{journal}{Prog. Part. Nucl. Phys.} \textbf{\bibinfo{volume}{108}},
  \bibinfo{pages}{103709} (\bibinfo{year}{2019}{\natexlab{c}}),
  \eprint{1811.04409}.

\bibitem[{\citenamefont{Montenegro and Torrieri}(2019)}]{Montenegro:2018bcf}
\bibinfo{author}{\bibfnamefont{D.}~\bibnamefont{Montenegro}} \bibnamefont{and}
  \bibinfo{author}{\bibfnamefont{G.}~\bibnamefont{Torrieri}},
  \bibinfo{journal}{Phys. Rev. D} \textbf{\bibinfo{volume}{100}},
  \bibinfo{pages}{056011} (\bibinfo{year}{2019}), \eprint{1807.02796}.

\bibitem[{\citenamefont{Hattori
  et~al.}(2019{\natexlab{b}})\citenamefont{Hattori, Hongo, Huang, Matsuo, and
  Taya}}]{Hattori:2019lfp}
\bibinfo{author}{\bibfnamefont{K.}~\bibnamefont{Hattori}},
  \bibinfo{author}{\bibfnamefont{M.}~\bibnamefont{Hongo}},
  \bibinfo{author}{\bibfnamefont{X.-G.} \bibnamefont{Huang}},
  \bibinfo{author}{\bibfnamefont{M.}~\bibnamefont{Matsuo}}, \bibnamefont{and}
  \bibinfo{author}{\bibfnamefont{H.}~\bibnamefont{Taya}},
  \bibinfo{journal}{Phys. Lett.} \textbf{\bibinfo{volume}{B795}},
  \bibinfo{pages}{100} (\bibinfo{year}{2019}{\natexlab{b}}),
  \eprint{1901.06615}.

\bibitem[{\citenamefont{Bhadury et~al.}(2021)\citenamefont{Bhadury, Florkowski,
  Jaiswal, Kumar, and Ryblewski}}]{Bhadury:2020puc}
\bibinfo{author}{\bibfnamefont{S.}~\bibnamefont{Bhadury}},
  \bibinfo{author}{\bibfnamefont{W.}~\bibnamefont{Florkowski}},
  \bibinfo{author}{\bibfnamefont{A.}~\bibnamefont{Jaiswal}},
  \bibinfo{author}{\bibfnamefont{A.}~\bibnamefont{Kumar}}, \bibnamefont{and}
  \bibinfo{author}{\bibfnamefont{R.}~\bibnamefont{Ryblewski}},
  \bibinfo{journal}{Phys. Lett. B} \textbf{\bibinfo{volume}{814}},
  \bibinfo{pages}{136096} (\bibinfo{year}{2021}), \eprint{2002.03937}.

\bibitem[{\citenamefont{Singh et~al.}(2021)\citenamefont{Singh, Sophys, and
  Ryblewski}}]{Singh:2020rht}
\bibinfo{author}{\bibfnamefont{R.}~\bibnamefont{Singh}},
  \bibinfo{author}{\bibfnamefont{G.}~\bibnamefont{Sophys}}, \bibnamefont{and}
  \bibinfo{author}{\bibfnamefont{R.}~\bibnamefont{Ryblewski}},
  \bibinfo{journal}{Phys. Rev. D} \textbf{\bibinfo{volume}{103}},
  \bibinfo{pages}{074024} (\bibinfo{year}{2021}), \eprint{2011.14907}.

\bibitem[{\citenamefont{Montenegro and Torrieri}(2020)}]{Montenegro:2020paq}
\bibinfo{author}{\bibfnamefont{D.}~\bibnamefont{Montenegro}} \bibnamefont{and}
  \bibinfo{author}{\bibfnamefont{G.}~\bibnamefont{Torrieri}},
  \bibinfo{journal}{Phys. Rev. D} \textbf{\bibinfo{volume}{102}},
  \bibinfo{pages}{036007} (\bibinfo{year}{2020}), \eprint{2004.10195}.

\bibitem[{\citenamefont{Gallegos et~al.}(2021)\citenamefont{Gallegos, G\"ursoy,
  and Yarom}}]{Gallegos:2021bzp}
\bibinfo{author}{\bibfnamefont{A.~D.} \bibnamefont{Gallegos}},
  \bibinfo{author}{\bibfnamefont{U.}~\bibnamefont{G\"ursoy}}, \bibnamefont{and}
  \bibinfo{author}{\bibfnamefont{A.}~\bibnamefont{Yarom}},
  \bibinfo{journal}{SciPost Phys.} \textbf{\bibinfo{volume}{11}},
  \bibinfo{pages}{041} (\bibinfo{year}{2021}), \eprint{2101.04759}.

\bibitem[{\citenamefont{Fukushima and Pu}(2021)}]{Fukushima:2020ucl}
\bibinfo{author}{\bibfnamefont{K.}~\bibnamefont{Fukushima}} \bibnamefont{and}
  \bibinfo{author}{\bibfnamefont{S.}~\bibnamefont{Pu}}, \bibinfo{journal}{Phys.
  Lett. B} \textbf{\bibinfo{volume}{817}}, \bibinfo{pages}{136346}
  (\bibinfo{year}{2021}), \eprint{2010.01608}.

\bibitem[{\citenamefont{Li et~al.}(2021)\citenamefont{Li, Stephanov, and
  Yee}}]{Li:2020eon}
\bibinfo{author}{\bibfnamefont{S.}~\bibnamefont{Li}},
  \bibinfo{author}{\bibfnamefont{M.~A.} \bibnamefont{Stephanov}},
  \bibnamefont{and} \bibinfo{author}{\bibfnamefont{H.-U.} \bibnamefont{Yee}},
  \bibinfo{journal}{Phys. Rev. Lett.} \textbf{\bibinfo{volume}{127}},
  \bibinfo{pages}{082302} (\bibinfo{year}{2021}), \eprint{2011.12318}.

\bibitem[{\citenamefont{Wang et~al.}(2021)\citenamefont{Wang, Fang, and
  Pu}}]{Wang:2021ngp}
\bibinfo{author}{\bibfnamefont{D.-L.} \bibnamefont{Wang}},
  \bibinfo{author}{\bibfnamefont{S.}~\bibnamefont{Fang}}, \bibnamefont{and}
  \bibinfo{author}{\bibfnamefont{S.}~\bibnamefont{Pu}}, \bibinfo{journal}{Phys.
  Rev. D} \textbf{\bibinfo{volume}{104}}, \bibinfo{pages}{114043}
  (\bibinfo{year}{2021}), \eprint{2107.11726}.

\bibitem[{\citenamefont{Hu}(2022)}]{Hu:2021pwh}
\bibinfo{author}{\bibfnamefont{J.}~\bibnamefont{Hu}}, \bibinfo{journal}{Phys.
  Rev. D} \textbf{\bibinfo{volume}{105}}, \bibinfo{pages}{076009}
  (\bibinfo{year}{2022}), \eprint{2111.03571}.

\bibitem[{\citenamefont{Hongo et~al.}(2021)\citenamefont{Hongo, Huang,
  Kaminski, Stephanov, and Yee}}]{Hongo:2021ona}
\bibinfo{author}{\bibfnamefont{M.}~\bibnamefont{Hongo}},
  \bibinfo{author}{\bibfnamefont{X.-G.} \bibnamefont{Huang}},
  \bibinfo{author}{\bibfnamefont{M.}~\bibnamefont{Kaminski}},
  \bibinfo{author}{\bibfnamefont{M.}~\bibnamefont{Stephanov}},
  \bibnamefont{and} \bibinfo{author}{\bibfnamefont{H.-U.} \bibnamefont{Yee}},
  \bibinfo{journal}{JHEP} \textbf{\bibinfo{volume}{11}}, \bibinfo{pages}{150}
  (\bibinfo{year}{2021}), \eprint{2107.14231}.

\bibitem[{\citenamefont{Daher et~al.}(2022)\citenamefont{Daher, Das,
  Florkowski, and Ryblewski}}]{Daher:2022xon}
\bibinfo{author}{\bibfnamefont{A.}~\bibnamefont{Daher}},
  \bibinfo{author}{\bibfnamefont{A.}~\bibnamefont{Das}},
  \bibinfo{author}{\bibfnamefont{W.}~\bibnamefont{Florkowski}},
  \bibnamefont{and} \bibinfo{author}{\bibfnamefont{R.}~\bibnamefont{Ryblewski}}
  (\bibinfo{year}{2022}), \eprint{2202.12609}.

\bibitem[{\citenamefont{Weickgenannt
  et~al.}(2022{\natexlab{a}})\citenamefont{Weickgenannt, Wagner, Speranza, and
  Rischke}}]{Weickgenannt:2022zxs}
\bibinfo{author}{\bibfnamefont{N.}~\bibnamefont{Weickgenannt}},
  \bibinfo{author}{\bibfnamefont{D.}~\bibnamefont{Wagner}},
  \bibinfo{author}{\bibfnamefont{E.}~\bibnamefont{Speranza}}, \bibnamefont{and}
  \bibinfo{author}{\bibfnamefont{D.~H.} \bibnamefont{Rischke}},
  \bibinfo{journal}{Phys. Rev. D} \textbf{\bibinfo{volume}{106}},
  \bibinfo{pages}{096014} (\bibinfo{year}{2022}{\natexlab{a}}),
  \eprint{2203.04766}.

\bibitem[{\citenamefont{Weickgenannt
  et~al.}(2022{\natexlab{b}})\citenamefont{Weickgenannt, Wagner, and
  Speranza}}]{Weickgenannt:2022jes}
\bibinfo{author}{\bibfnamefont{N.}~\bibnamefont{Weickgenannt}},
  \bibinfo{author}{\bibfnamefont{D.}~\bibnamefont{Wagner}}, \bibnamefont{and}
  \bibinfo{author}{\bibfnamefont{E.}~\bibnamefont{Speranza}},
  \bibinfo{journal}{Phys. Rev. D} \textbf{\bibinfo{volume}{105}},
  \bibinfo{pages}{116026} (\bibinfo{year}{2022}{\natexlab{b}}),
  \eprint{2204.01797}.

\bibitem[{\citenamefont{Gallegos et~al.}(2022)\citenamefont{Gallegos, Gursoy,
  and Yarom}}]{Gallegos:2022jow}
\bibinfo{author}{\bibfnamefont{A.~D.} \bibnamefont{Gallegos}},
  \bibinfo{author}{\bibfnamefont{U.}~\bibnamefont{Gursoy}}, \bibnamefont{and}
  \bibinfo{author}{\bibfnamefont{A.}~\bibnamefont{Yarom}}
  (\bibinfo{year}{2022}), \eprint{2203.05044}.

\bibitem[{\citenamefont{Cao et~al.}(2022)\citenamefont{Cao, Hattori, Hongo,
  Huang, and Taya}}]{Cao:2022aku}
\bibinfo{author}{\bibfnamefont{Z.}~\bibnamefont{Cao}},
  \bibinfo{author}{\bibfnamefont{K.}~\bibnamefont{Hattori}},
  \bibinfo{author}{\bibfnamefont{M.}~\bibnamefont{Hongo}},
  \bibinfo{author}{\bibfnamefont{X.-G.} \bibnamefont{Huang}}, \bibnamefont{and}
  \bibinfo{author}{\bibfnamefont{H.}~\bibnamefont{Taya}}
  (\bibinfo{year}{2022}), \eprint{2205.08051}.

\bibitem[{\citenamefont{Weickgenannt
  et~al.}(2022{\natexlab{c}})\citenamefont{Weickgenannt, Wagner, Speranza, and
  Rischke}}]{Weickgenannt:2022qvh}
\bibinfo{author}{\bibfnamefont{N.}~\bibnamefont{Weickgenannt}},
  \bibinfo{author}{\bibfnamefont{D.}~\bibnamefont{Wagner}},
  \bibinfo{author}{\bibfnamefont{E.}~\bibnamefont{Speranza}}, \bibnamefont{and}
  \bibinfo{author}{\bibfnamefont{D.~H.} \bibnamefont{Rischke}},
  \bibinfo{journal}{Phys. Rev. D} \textbf{\bibinfo{volume}{106}},
  \bibinfo{pages}{L091901} (\bibinfo{year}{2022}{\natexlab{c}}),
  \eprint{2208.01955}.

\bibitem[{\citenamefont{Biswas et~al.}(2023)\citenamefont{Biswas, Daher, Das,
  Florkowski, and Ryblewski}}]{Biswas:2023qsw}
\bibinfo{author}{\bibfnamefont{R.}~\bibnamefont{Biswas}},
  \bibinfo{author}{\bibfnamefont{A.}~\bibnamefont{Daher}},
  \bibinfo{author}{\bibfnamefont{A.}~\bibnamefont{Das}},
  \bibinfo{author}{\bibfnamefont{W.}~\bibnamefont{Florkowski}},
  \bibnamefont{and} \bibinfo{author}{\bibfnamefont{R.}~\bibnamefont{Ryblewski}}
  (\bibinfo{year}{2023}), \eprint{2304.01009}.

\bibitem[{\citenamefont{Weickgenannt}(2023)}]{Weickgenannt:2023btk}
\bibinfo{author}{\bibfnamefont{N.}~\bibnamefont{Weickgenannt}},
  \bibinfo{journal}{Phys. Rev. D} \textbf{\bibinfo{volume}{108}},
  \bibinfo{pages}{076011} (\bibinfo{year}{2023}), \eprint{2307.13561}.

\bibitem[{\citenamefont{Weickgenannt and Blaizot}(2024)}]{Weickgenannt:2023bss}
\bibinfo{author}{\bibfnamefont{N.}~\bibnamefont{Weickgenannt}}
  \bibnamefont{and} \bibinfo{author}{\bibfnamefont{J.-P.}
  \bibnamefont{Blaizot}}, \bibinfo{journal}{Phys. Rev. D}
  \textbf{\bibinfo{volume}{109}}, \bibinfo{pages}{056019}
  (\bibinfo{year}{2024}), \eprint{2312.05917}.

\bibitem[{\citenamefont{Daher et~al.}(2024)\citenamefont{Daher, Florkowski,
  Ryblewski, and Taghinavaz}}]{Daher:2024bah}
\bibinfo{author}{\bibfnamefont{A.}~\bibnamefont{Daher}},
  \bibinfo{author}{\bibfnamefont{W.}~\bibnamefont{Florkowski}},
  \bibinfo{author}{\bibfnamefont{R.}~\bibnamefont{Ryblewski}},
  \bibnamefont{and}
  \bibinfo{author}{\bibfnamefont{F.}~\bibnamefont{Taghinavaz}},
  \bibinfo{journal}{Phys. Rev. D} \textbf{\bibinfo{volume}{109}},
  \bibinfo{pages}{114001} (\bibinfo{year}{2024}), \eprint{2403.04711}.

\bibitem[{\citenamefont{Wagner et~al.}(2024)\citenamefont{Wagner, Shokri, and
  Rischke}}]{Wagner:2024fhf}
\bibinfo{author}{\bibfnamefont{D.}~\bibnamefont{Wagner}},
  \bibinfo{author}{\bibfnamefont{M.}~\bibnamefont{Shokri}}, \bibnamefont{and}
  \bibinfo{author}{\bibfnamefont{D.~H.} \bibnamefont{Rischke}}
  (\bibinfo{year}{2024}), \eprint{2405.00533}.

\bibitem[{\citenamefont{Drogosz et~al.}(2024)\citenamefont{Drogosz, Florkowski,
  and Hontarenko}}]{Drogosz:2024gzv}
\bibinfo{author}{\bibfnamefont{Z.}~\bibnamefont{Drogosz}},
  \bibinfo{author}{\bibfnamefont{W.}~\bibnamefont{Florkowski}},
  \bibnamefont{and}
  \bibinfo{author}{\bibfnamefont{M.}~\bibnamefont{Hontarenko}}
  (\bibinfo{year}{2024}), \eprint{2408.03106}.

\bibitem[{\citenamefont{Wagner}(2024)}]{Wagner:2024fry}
\bibinfo{author}{\bibfnamefont{D.}~\bibnamefont{Wagner}}
  (\bibinfo{year}{2024}), \eprint{2409.07143}.

\bibitem[{\citenamefont{Bemfica et~al.}(2018)\citenamefont{Bemfica, Disconzi,
  and Noronha}}]{Bemfica:2017wps}
\bibinfo{author}{\bibfnamefont{F.~S.} \bibnamefont{Bemfica}},
  \bibinfo{author}{\bibfnamefont{M.~M.} \bibnamefont{Disconzi}},
  \bibnamefont{and} \bibinfo{author}{\bibfnamefont{J.}~\bibnamefont{Noronha}},
  \bibinfo{journal}{Phys. Rev. D} \textbf{\bibinfo{volume}{98}},
  \bibinfo{pages}{104064} (\bibinfo{year}{2018}), \eprint{1708.06255}.

\bibitem[{\citenamefont{Bemfica et~al.}(2019)\citenamefont{Bemfica, Disconzi,
  and Noronha}}]{Bemfica:2019knx}
\bibinfo{author}{\bibfnamefont{F.~S.} \bibnamefont{Bemfica}},
  \bibinfo{author}{\bibfnamefont{M.~M.} \bibnamefont{Disconzi}},
  \bibnamefont{and} \bibinfo{author}{\bibfnamefont{J.}~\bibnamefont{Noronha}},
  \bibinfo{journal}{Phys. Rev. D} \textbf{\bibinfo{volume}{100}},
  \bibinfo{pages}{104020} (\bibinfo{year}{2019}), \eprint{1907.12695}.

\bibitem[{\citenamefont{Kovtun}(2019)}]{Kovtun:2019hdm}
\bibinfo{author}{\bibfnamefont{P.}~\bibnamefont{Kovtun}},
  \bibinfo{journal}{JHEP} \textbf{\bibinfo{volume}{10}}, \bibinfo{pages}{034}
  (\bibinfo{year}{2019}), \eprint{1907.08191}.

\bibitem[{\citenamefont{Bemfica et~al.}(2020)\citenamefont{Bemfica, Disconzi,
  and Noronha}}]{Bemfica:2020zjp}
\bibinfo{author}{\bibfnamefont{F.~S.} \bibnamefont{Bemfica}},
  \bibinfo{author}{\bibfnamefont{M.~M.} \bibnamefont{Disconzi}},
  \bibnamefont{and} \bibinfo{author}{\bibfnamefont{J.}~\bibnamefont{Noronha}}
  (\bibinfo{year}{2020}), \eprint{2009.11388}.

\bibitem[{\citenamefont{Hoult and Kovtun}(2020)}]{Hoult:2020eho}
\bibinfo{author}{\bibfnamefont{R.~E.} \bibnamefont{Hoult}} \bibnamefont{and}
  \bibinfo{author}{\bibfnamefont{P.}~\bibnamefont{Kovtun}},
  \bibinfo{journal}{JHEP} \textbf{\bibinfo{volume}{06}}, \bibinfo{pages}{067}
  (\bibinfo{year}{2020}), \eprint{2004.04102}.

\bibitem[{\citenamefont{Abboud et~al.}(2023)\citenamefont{Abboud, Speranza, and
  Noronha}}]{Abboud:2023hos}
\bibinfo{author}{\bibfnamefont{N.}~\bibnamefont{Abboud}},
  \bibinfo{author}{\bibfnamefont{E.}~\bibnamefont{Speranza}}, \bibnamefont{and}
  \bibinfo{author}{\bibfnamefont{J.}~\bibnamefont{Noronha}}
  (\bibinfo{year}{2023}), \eprint{2308.02928}.

\bibitem[{\citenamefont{Becattini
  et~al.}(2013{\natexlab{b}})\citenamefont{Becattini, Csernai, and
  Wang}}]{Becattini:2013vja}
\bibinfo{author}{\bibfnamefont{F.}~\bibnamefont{Becattini}},
  \bibinfo{author}{\bibfnamefont{L.}~\bibnamefont{Csernai}}, \bibnamefont{and}
  \bibinfo{author}{\bibfnamefont{D.~J.} \bibnamefont{Wang}},
  \bibinfo{journal}{Phys. Rev.} \textbf{\bibinfo{volume}{C88}},
  \bibinfo{pages}{034905} (\bibinfo{year}{2013}{\natexlab{b}}),
  \bibinfo{note}{[Erratum: Phys. Rev.C93,no.6,069901(2016)]},
  \eprint{1304.4427}.

\bibitem[{\citenamefont{Becattini et~al.}(2015)\citenamefont{Becattini,
  Inghirami, Rolando, Beraudo, Del~Zanna, De~Pace, Nardi, Pagliara, and
  Chandra}}]{Becattini:2015ska}
\bibinfo{author}{\bibfnamefont{F.}~\bibnamefont{Becattini}},
  \bibinfo{author}{\bibfnamefont{G.}~\bibnamefont{Inghirami}},
  \bibinfo{author}{\bibfnamefont{V.}~\bibnamefont{Rolando}},
  \bibinfo{author}{\bibfnamefont{A.}~\bibnamefont{Beraudo}},
  \bibinfo{author}{\bibfnamefont{L.}~\bibnamefont{Del~Zanna}},
  \bibinfo{author}{\bibfnamefont{A.}~\bibnamefont{De~Pace}},
  \bibinfo{author}{\bibfnamefont{M.}~\bibnamefont{Nardi}},
  \bibinfo{author}{\bibfnamefont{G.}~\bibnamefont{Pagliara}}, \bibnamefont{and}
  \bibinfo{author}{\bibfnamefont{V.}~\bibnamefont{Chandra}},
  \bibinfo{journal}{Eur. Phys. J.} \textbf{\bibinfo{volume}{C75}},
  \bibinfo{pages}{406} (\bibinfo{year}{2015}), \bibinfo{note}{[Erratum: Eur.
  Phys. J.C78,no.5,354(2018)]}, \eprint{1501.04468}.

\bibitem[{\citenamefont{Becattini et~al.}(2017)\citenamefont{Becattini,
  Karpenko, Lisa, Upsal, and Voloshin}}]{Becattini:2016gvu}
\bibinfo{author}{\bibfnamefont{F.}~\bibnamefont{Becattini}},
  \bibinfo{author}{\bibfnamefont{I.}~\bibnamefont{Karpenko}},
  \bibinfo{author}{\bibfnamefont{M.}~\bibnamefont{Lisa}},
  \bibinfo{author}{\bibfnamefont{I.}~\bibnamefont{Upsal}}, \bibnamefont{and}
  \bibinfo{author}{\bibfnamefont{S.}~\bibnamefont{Voloshin}},
  \bibinfo{journal}{Phys. Rev.} \textbf{\bibinfo{volume}{C95}},
  \bibinfo{pages}{054902} (\bibinfo{year}{2017}), \eprint{1610.02506}.

\bibitem[{\citenamefont{Karpenko and Becattini}(2017)}]{Karpenko:2016jyx}
\bibinfo{author}{\bibfnamefont{I.}~\bibnamefont{Karpenko}} \bibnamefont{and}
  \bibinfo{author}{\bibfnamefont{F.}~\bibnamefont{Becattini}},
  \bibinfo{journal}{Eur. Phys. J.} \textbf{\bibinfo{volume}{C77}},
  \bibinfo{pages}{213} (\bibinfo{year}{2017}), \eprint{1610.04717}.

\bibitem[{\citenamefont{Pang et~al.}(2016)\citenamefont{Pang, Petersen, Wang,
  and Wang}}]{Pang:2016igs}
\bibinfo{author}{\bibfnamefont{L.-G.} \bibnamefont{Pang}},
  \bibinfo{author}{\bibfnamefont{H.}~\bibnamefont{Petersen}},
  \bibinfo{author}{\bibfnamefont{Q.}~\bibnamefont{Wang}}, \bibnamefont{and}
  \bibinfo{author}{\bibfnamefont{X.-N.} \bibnamefont{Wang}},
  \bibinfo{journal}{Phys. Rev. Lett.} \textbf{\bibinfo{volume}{117}},
  \bibinfo{pages}{192301} (\bibinfo{year}{2016}), \eprint{1605.04024}.

\bibitem[{\citenamefont{Xie et~al.}(2017)\citenamefont{Xie, Wang, and
  Csernai}}]{Xie:2017upb}
\bibinfo{author}{\bibfnamefont{Y.}~\bibnamefont{Xie}},
  \bibinfo{author}{\bibfnamefont{D.}~\bibnamefont{Wang}}, \bibnamefont{and}
  \bibinfo{author}{\bibfnamefont{L.~P.} \bibnamefont{Csernai}},
  \bibinfo{journal}{Phys. Rev.} \textbf{\bibinfo{volume}{C95}},
  \bibinfo{pages}{031901} (\bibinfo{year}{2017}), \eprint{1703.03770}.

\bibitem[{\citenamefont{Zubarev}(1966)}]{zubarev1966statistical}
\bibinfo{author}{\bibfnamefont{D.}~\bibnamefont{Zubarev}}, in
  \emph{\bibinfo{booktitle}{Soviet Physics Doklady}} (\bibinfo{year}{1966}),
  vol.~\bibinfo{volume}{10}, p. \bibinfo{pages}{850}.

\bibitem[{\citenamefont{Alzhrani et~al.}(2022)\citenamefont{Alzhrani, Ryu, and
  Shen}}]{Alzhrani:2022dpi}
\bibinfo{author}{\bibfnamefont{S.}~\bibnamefont{Alzhrani}},
  \bibinfo{author}{\bibfnamefont{S.}~\bibnamefont{Ryu}}, \bibnamefont{and}
  \bibinfo{author}{\bibfnamefont{C.}~\bibnamefont{Shen}},
  \bibinfo{journal}{Phys. Rev. C} \textbf{\bibinfo{volume}{106}},
  \bibinfo{pages}{014905} (\bibinfo{year}{2022}), \eprint{2203.15718}.

\bibitem[{\citenamefont{Jiang et~al.}(2023)\citenamefont{Jiang, Wu, Cao, and
  Zhang}}]{Jiang:2023vxp}
\bibinfo{author}{\bibfnamefont{Z.-F.} \bibnamefont{Jiang}},
  \bibinfo{author}{\bibfnamefont{X.-Y.} \bibnamefont{Wu}},
  \bibinfo{author}{\bibfnamefont{S.}~\bibnamefont{Cao}}, \bibnamefont{and}
  \bibinfo{author}{\bibfnamefont{B.-W.} \bibnamefont{Zhang}}
  (\bibinfo{year}{2023}), \eprint{2307.04257}.

\bibitem[{\citenamefont{Palermo et~al.}(2024)\citenamefont{Palermo, Grossi,
  Karpenko, and Becattini}}]{Palermo:2024tza}
\bibinfo{author}{\bibfnamefont{A.}~\bibnamefont{Palermo}},
  \bibinfo{author}{\bibfnamefont{E.}~\bibnamefont{Grossi}},
  \bibinfo{author}{\bibfnamefont{I.}~\bibnamefont{Karpenko}}, \bibnamefont{and}
  \bibinfo{author}{\bibfnamefont{F.}~\bibnamefont{Becattini}}
  (\bibinfo{year}{2024}), \eprint{2404.14295}.

\bibitem[{\citenamefont{Hehl}(1976)}]{Hehl:1976vr}
\bibinfo{author}{\bibfnamefont{F.~W.} \bibnamefont{Hehl}},
  \bibinfo{journal}{Rept. Math. Phys.} \textbf{\bibinfo{volume}{9}},
  \bibinfo{pages}{55} (\bibinfo{year}{1976}).

\bibitem[{\citenamefont{Becattini et~al.}(2019)\citenamefont{Becattini,
  Florkowski, and Speranza}}]{Becattini:2018duy}
\bibinfo{author}{\bibfnamefont{F.}~\bibnamefont{Becattini}},
  \bibinfo{author}{\bibfnamefont{W.}~\bibnamefont{Florkowski}},
  \bibnamefont{and} \bibinfo{author}{\bibfnamefont{E.}~\bibnamefont{Speranza}},
  \bibinfo{journal}{Phys. Lett.} \textbf{\bibinfo{volume}{B789}},
  \bibinfo{pages}{419} (\bibinfo{year}{2019}), \eprint{1807.10994}.

\bibitem[{\citenamefont{Speranza and Weickgenannt}(2021)}]{Speranza:2020ilk}
\bibinfo{author}{\bibfnamefont{E.}~\bibnamefont{Speranza}} \bibnamefont{and}
  \bibinfo{author}{\bibfnamefont{N.}~\bibnamefont{Weickgenannt}},
  \bibinfo{journal}{Eur. Phys. J. A} \textbf{\bibinfo{volume}{57}},
  \bibinfo{pages}{155} (\bibinfo{year}{2021}), \eprint{2007.00138}.

\bibitem[{\citenamefont{Hilgevoord and Wouthuysen}(1963)}]{HILGEVOORD19631}
\bibinfo{author}{\bibfnamefont{J.}~\bibnamefont{Hilgevoord}} \bibnamefont{and}
  \bibinfo{author}{\bibfnamefont{S.}~\bibnamefont{Wouthuysen}},
  \bibinfo{journal}{Nuclear Physics} \textbf{\bibinfo{volume}{40}},
  \bibinfo{pages}{1 } (\bibinfo{year}{1963}), ISSN \bibinfo{issn}{0029-5582}.

\bibitem[{\citenamefont{Vollhardt and Wolfle}(2013)}]{vollhardt2013superfluid}
\bibinfo{author}{\bibfnamefont{D.}~\bibnamefont{Vollhardt}} \bibnamefont{and}
  \bibinfo{author}{\bibfnamefont{P.}~\bibnamefont{Wolfle}},
  \emph{\bibinfo{title}{The superfluid phases of helium 3}}
  (\bibinfo{publisher}{Courier Corporation}, \bibinfo{year}{2013}).

\end{thebibliography}

\end{document}